\documentclass[psfig]{mn2e}
\usepackage{graphicx}

\begin{document}

\title[Spectroscopy of the Starburst Galaxy NGC 4861]
 {Integral Field Spectroscopy of the Cometary Starburst Galaxy NGC 4861}
\author[N. Roche, J.M. V\'ilchez, J. Iglesias-P\'aramo, P. Papaderos, S.F. S\'anchez, C. Kehrig, S. Duarte Puertas]
 {Nathan Roche$^1$\thanks{nroche@iaa.es; nathanroche@mac.com},
 Jos\'e M. V\'ilchez$^{1}$,  Jorge Iglesias-P\'aramo$^{1}$, Polychronis Papaderos$^{2}$\\\\
 \noindent {\rm \LARGE Sebastian F. S\'anchez$^{3}$}, {\LARGE  \rm  Carolina Kehrig$^{1}$}, {\rm \LARGE Salvador Duarte Puertas$^{4,5,1}$}\\\\
$^1$ Instituto de Astrof\'isica de Andaluc\'ia-CSIC, Glorieta de la Astronomia, Granada 18008,
Spain.\\
$^2$ Instituto de Astrof\'isica e Ci\^encias do Espa\c co, 
Universidade do Porto, CAUP, Rua das Estrelas, 4150-762 Porto, Portugal.\\
$^3$ Instituto de Astronom\'ia, Universidad Nacional Autonoma de M\'exico, AP 70-264, CDMX 04510, Mexico\\
$^4$ Departamento de F\'isica Te\'orica y del Cosmos, Universidad de Granada, Campus Fuente Nueva, Edificio Mecenas,\\ Granada 18071, Spain.\\
$^5$ D\'epartement de Physique, de G\'enie Physique et d'Optique, Universit\'e Laval, and Centre de Recherche en Astrophysique du Qu\'ebec\\ (CRAQ), Qu\'ebec, QC, G1V 0A6, Canada\\}

\bibliographystyle{unsrt} \bibliographystyle{unsrt}

\date{18 April 2023}

\pagerange{\pageref{firstpage}--\pageref{lastpage}} \pubyear{2023}

\maketitle
 
\label{firstpage}

\begin{abstract} 

Using the PMAS Integral Field Unit on the Calar Alto 3.5m telescope we observed the southern component (Markarian 59) of the `cometary' starburst galaxy NGC 4861. Mrk 59 is centred on a giant nebula and concentration of stars 1 kpc in diameter. Strong $\rm H\alpha$ emission points to a star-formation rate (SFR) at least 0.47 $\rm M_{\odot}yr^{-1}$. Mrk 59 has a very high [OIII]$\rm\lambda5007/H\beta$ ratio, reaching 7.35 in the central nebula, with a second peak at a star-forming hotspot further north. Fast outflows are not detected but nebular motion and galaxy rotation produce relative velocities up to 40 km $\rm s^{-1}$. Spectral analysis of different regions with `Fitting Analysis using Differential evolution Optimisation'  (FADO) finds that the stars in the central and `spur' nebulae are very young, $\rm \leq125~Myr$ with a large $\rm <10~Myr$ contribution. Older stars ($\rm \sim 1~Gyr$) make up the northern disk component, while the other regions show mixtures of 1 Gyr age with very young stars. This and the high specific SFR $\rm\sim 3.5~Gyr^{-1}$ imply a bimodal star formation history, with Mrk 59 formed in ongoing starbursts fuelled by a huge gas inflow, turning the galaxy into an asymmetric `green pea' or blue compact dwarf. We map the HeII$\lambda4686$ emission, and identify a broad component from the central nebula, consistent with the emission of $\sim 300$ Wolf-Rayet stars. About a third of the HeII$\lambda$4686 flux is a narrow line emitted from a more extended area covering the central and spur nebulae, and may have a different origin.

    \end{abstract}

\begin{keywords}
 galaxies: individual: NGC 4861: Markarian 59, galaxies: evolution, galaxies: starburst, stars: Wolf-Rayet
 \end{keywords}

\section{Introduction}

The galaxy NGC 4861 was listed in the Arp Atlas of Peculiar Galaxies (Arp 1966) as number 266, in the category of a `galaxy with irregular clumps', and was noted to have a `bright knot at S end'. Wakamatsu et al. (1979) described the galaxy as `resembling the shape of a comet'. The cometary `head', at the southern end of NGC 4861 has a much higher surface brightness than the rest of the galaxy (a relatively faint `tail'), and has been separately catalogued as Markarian 59 (Mrk 59).  Wakamatsu et al. (1979) performed spectroscopy of Mrk 59 and found it to be centred on a  giant HII region, a nebula undergoing intense star-formation, as evident from its very strong emission lines, especially $\rm H\alpha$ and $\rm [OIII]\lambda5007$ for which they estimated equivalent widths (EWs) $\rm >1000\AA$. They also noted the galaxy lacked close neighbours  and so its burst of star-formation might not have been triggered by an interaction.

 High-resolution spectroscopy of the HII region (Dinerstein \& Shields 1986) gave a redshift $z=0.0027$ and revealed a fainter emission line of ionized helium, HeII$\lambda4686$, associated with a `bump' in the continuum. This is the signature of Wolf-Rayet (WR) stars, which are very hot, massive and short-lived objects -- their presence in a galaxy indicates that a significant fraction of the stars formed in the past few Myr. Conti (1991) therefore included NGC 4861 in a catalog of `Wolf-Rayet galaxies', where the criterion was the detection of a broad  HeII$\lambda$4686 emission feature (with full width at half-maximum, $\rm FWHM\geq 10\AA$), as emitted directly from WR stars.
 
 Dottori et al. (1994) performed CCD imaging of NGC 4861 in narrow and broad bands, and found, in addition to the giant HII region, a chain of much smaller $\rm H\alpha$ emitting `hotspots' extending northwards along the low surface brightness `tail'. Barth et al. (1994) used this data to map the galaxy in detail, listing 28 line-emitting regions. They found very high $\rm [OIII]\lambda5007/H\beta$ (excitation) ratios at the giant nebula (reaching $\geq 6$), which tended to decrease with distance from it (northwards). They suggested this was related to a stellar age gradient, with star-formation having propagated southwards along the chain of hotspots to reach the giant nebula, at hundreds of km $\rm s^{-1}$. Schaerer, Contini \& Pindao (1999) included NGC 4861 in a new catalog of WR galaxies with broad HeII emission. 
 
 Noeske et al. (2000) described the galaxy as a prototypical `Cometary Blue Compact Dwarf', and performed deep long-slit spectroscopy. Comparing with spectral synthesis models they concluded the starbursting giant nebula contains mostly very young ($\sim 4$ Myr) stars while the remainder of the galaxy is a mixture of young stars ($\sim$ 10--25 Myr) and an underlying older population of age $\sim 2$ Gyr. At the centre of the giant nebula they estimate a metallicity $\rm 12+log(O/H)\simeq 8.0$, and detect both WR features and narrow (nebular) HeII$\lambda$4686 emission.
 
 Thuan, Hibbard \& L\'evrier (2004) performed HI observations with the Very Large Array (VLA). They found the main body of NGC 4861 to be an almost edge-on ($82^{\circ}$) disk galaxy undergoing regular rotation, approaching at the southern end, with a maximum rotation velocity 52--54 km $\rm s^{-1}$. The giant nebula, at the south, is also approaching. However, while the disk shows a near linear rotation curve over its extent, the Mrk 59 component has a more uniform radial velocity, with a discontinuity of $\sim -20$ km $\rm s^{-1}$  from the southern edge of the disk. The total HI mass of NGC 4861 is estimated as  $\rm1.1\times 10^9M_{\odot}$.	 
 
 Fernandes et al. (2004) examined the optical spectra of  WR galaxies, including NGC 4861, which 
they estimated to contain $586\pm 280$ WR stars, mostly of type WNL (522), again with a best-fit starburst age of 4.0 Myr.
 Van Eymeren et al. (2007) performed Echelle spectroscopy in $\rm H\alpha$ with several N--S slit positions, and find `ionized shells and filaments', with a blueshifted region $\rm (SGS4_{blue})$ west of the giant nebula, which could be `part of an expanding shell', and a slightly redshifted `outflow' region about 40 arcsec north of the nebula. Van Eymeren et al. (2009), with Fabry-Perot  $\rm H\alpha$ combined with HI data, saw galaxy rotation with a radial velocity 46 km $\rm s^{-1}$ plus evidence for outflows with both blueshift and redshift components. Two are near to and north and south of the giant HII region and may be part of `supergiant shell' SGS4. However the velocities measured for outflows ($\simeq 30$ km $\rm s^{-1}$) are insufficient for the gas to escape from the galaxy. 
 
 Karthick et al. (2014) examined NGC 4861 as part of a study of WR galaxies. They estimated the metallicity as $\rm 12+log (O/H)=7.95\pm 0.05$, the age of the main starburst as 4.6 Myr and the star-formation rate (SFR) as $0.48\pm 0.04  \rm~ M_{\odot}yr^{-1}$. They depict its WR `blue bump'  as mildly broadened HeII$\lambda$4686 forming a group of four emission lines with  [Fe III]$\lambda4658$ and $\rm [Ar IV]\lambda\lambda4711,4740$. By fitting narrow and broad Gaussians they separate narrow and broad components of HeII, and estimate the total number of WR stars, dividing into two types as $\rm WNL=225\pm35$, $\rm WCE=67\pm 30$. 
  
 We also make comparisons with the extreme starburst galaxies known as `Green Peas' (Cardamone et al. 2009). Most examples are more massive and distant than NGC 4861 but have similar spectra:
  $\rm H\alpha$ and [OIII]$\lambda$5007 equivalent widths (EWs) of several hundreds of Angstroms, high $\rm [OIII]\lambda5007/H\beta$ ratios ($>5$) and low metallicity. They have very high specific Star Formation Rates (sSFRs) of 1--100 $\rm Gyr^{-1}$ (Izotov, Guseva \& Thuan 2011) but little or no AGN contribution. They may show spectral signatures of WR stars (Amor\'in et al. 2012), and strong emission in Lyman-$\alpha$ and shorter UV wavelengths, resembling formative galaxies at $z>3$ (Henry et al. 2015, Izotov et al. 2021). Green Peas may also contain older ($>1$ Gyr) stellar populations but these can be difficult to distinguish from diffuse nebular continuum which can make up $>10\%$ of their flux (Clarke et al. 2021; Fernandez et al. 2022).
  Mrk 59/NGC 4861, along with the cometary Blue Compact Dwarf Mrk 71/NGC 2366 (Micheva et al. 2017), are considered the nearest green pea `analogs'. On account of their small mass and low redshift both might also go in to the class of `blueberry' dwarf starbursts (Liu et al. 2022).
 
 All three areas, of star-formation history, kinematics, and the HeII/WR emission, can be investigated with more detail with the 2D spatial as well as spectral resolution
 provided by integral field spectroscopy  (IFS).
 In this study of NGC 4861/Mrk 59 we perform IFS at optical wavelengths with the Potsdam Multi Aperture Spectrophotometer (PMAS-PPAK) on Calar Alto, the same set-up as in the Calar Alto Legacy Integral Field Area (CALIFA) survey of hundreds of nearby galaxies (e.g. S\'anchez et al 2012, Garc\'ia-Benito et al. 2015). We also make use of Hubble Space Telescope data and  spectral synthesis models to fit the spectra.
  \begin{table}
\begin{tabular}{lc}
\hline
Name & Markarian 59\\
Names for whole system &
NGC 4861, Arp 266\\
   &       UGC 08098\\
R.A. & 12:59:00.3\\
Declination & +34:50:43\\
Spectroscopic Redshift & 0.00264 \\
Recession velocity & 792 km $\rm s^{-1}$ \\ 
Distance (adopted) & 15.9 Mpc \\
Scale  & 77 pc $\rm arcsec^{-1}$ \\
Distance modulus & 31.01 mag \\
 \hline
\end{tabular}
\caption{Some quantities (J2000, from IPAC NED and used in this paper) for Markarian 59, aka the southern component of NGC 4861.}

\end{table}

From IPAC NED (ned.ipac.caltech.edu), the J2000 co-ordinates of NGC 4861 (the disk galaxy centre) are R.A. 12h59m02.34s and Dec +34d51m34.0s,
 and the heliocentric recession velocity is 835 km $\rm s^{-1}$ with redshift $z=0.00279$, while the velocity relative to the CMB 3K background is given as 1079 km $\rm s^{-1}$, which would correspond to $z=0.003599$ or 15.9 Mpc.
 
 For the southern (giant nebula) component Markarian 59, NED lists co-ordinates
 R.A. 12h59m00.288s,	+34d50m42.47s (57 arcsec from the NGC 4861 centre on position angle $206^{\circ}$), and a heliocentric velocity slightly lower at 792 km $\rm s^{-1}$ ($z=0.00264$), with the velocity relative to the CMB 3K background likewise reduced to 1036 km $\rm s^{-1}$. This difference in radial velocity is due to the rotation of NGC 4861 and is consistent with the rotation curve measured in HI (Eymeren et al. 2007).
 
We adopt here the `cosmology corrected' distance for the centre of the NGC 4861 disk, 
$15.9 (\pm 1.1)$ Mpc, in setting a distance modulus 31.01 mag and angular scale 77 parsec $\rm arcsec^{-1}$ (again from NED, which assumed $H_0=67.8$ km $\rm s^{-1}Mpc^{-1}$, $\Omega_m=0.308$, 
$\Omega_\Lambda=0.692$) in the subsequent analysis (Table 1). 
Mrk 59 is listed in the Sloan Digital Sky Survey with magnitudes (g,r,i,z)=(13.88, 14.16, 14.90, 14.94) giving $M_g=-17.13$ and extremely blue colours.

This paper is organised as follows: in section 2 we describe the observational data, in section 3 the present-day appearance and activity of the galaxy, in section 4 the kinematics map from radial velocities, in section 5 we attempt to reconstruct the star-formation history of different regions, section 6 focusses on the Helium II emission line and finally section 7 is a summary and discussion.

 \section{Observations} 
 \subsection{PMAS Integral Field Unit Data}
 Galaxy NGC 4861 was observed using the PMAS integral field unit on the 3.5m aperture telescope at Calar Alto, Spain (Roth et al. 2005).  As in the CALIFA survey, PMAS was operated with the fibre bundle PPAK to give a larger ($>$1 arcmin) field-of-view. Our principal dataset for this paper is a data cube 
 centred on the southern part of galaxy NGC 4861, a few arcsec north of the giant nebula. The field of view  covers the southern (Mrk 59) component and only a fractional (but still useful) part of the disk galaxy to the north.  Our observations taken on 9 March 2019 followed the standard CALIFA set-up with three
 dithered exposures of 600 seconds, totalling 1800 seconds on target. Airmass was 1.08 to 1.17. Sky exposures ($3\times 100$ sec)  were obtained nearby and immediately afterwards. Standard star BD+33d2642 was used for spectrophotometric calibration. 
  
  The PPAK unit contains 331 densely packed optical fibres arranged in a hexagonal area of $74\times 65$ arcsec, to sample an astronomical object at 2.7 arcsec per fibre, with a filling factor of 65\% due to gaps in between the fibres. Our observations used the V500 grating and covered $\lambda=3750$--7500$\rm \AA$ with a spectral resolution of $6.0\rm \AA$ (FWHM) and dispersion $2\rm \AA$ $\rm pixel^{-1}$. The data were corrected for Galactic dust extinction of $A_V=0.029$ mag (from IPAC NED; Schlafly \& Finkbeiner 2011) and flux calibrated in units of $10^{-16}$ erg $\rm cm^{-2}s^{-1}$. Our data cubes are in the format of 1 arcsec spatial pixels (spaxels) arranged in a $78\times 73$ image,  While this makes up 5694 pixels, the approximately hexagonal area observed by PPAK only covers 4078 of these. 
 
 \subsection{HST Imaging Data}
 We also examine two Hubble Space Telescope (HST) images of NGC 4861, downloaded from the MAST public archive (mast.stsci.edu), for much higher spatial resolution views of the stellar and nebular components. These were taken with Wide Field Camera 3 in the broad $I$-band F814W, and in the narrow band F680N (FWHM $150.8\rm\AA$) which contains the $\rm H\alpha$ line. The first has the Observation ID hst\_12497\_02\_wfc3\_uvis\_f814w\_ibse02 and the second the ID  hst\_12497\_02\_wfc3\_uvis\_f658n\_ibse02 (both are from the proposal 12497 of  Sungryong Hong).
 
 \subsection{Data Processing}
 
 Beginning with the calibrated $xy\lambda$ PMAS data cube, a  processed data cube was then  generated with  all the spectra de-redshifted and rebinned in $1\rm \AA$ pixels. The radial velocity of each spaxel  relative to the systematic recession velocity of 790 km $\rm s^{-1}$ was measured from strong emission lines.
The spaxels in outer, low surface-brightness regions were binned into larger areas using a Voronoi tesselation to give a more useful signal/noise ratio, while over the higher surface brightness body of the galaxy the data is kept in the original 1 $\rm arcsec^2$ spaxels. This moderate binning reduced the number of different spectra in the processed data cube from 4078 to 1312. 
 
The next step was the fitting of composite stellar populations to the spectra in each of these 1312 spaxels (or  multi-spaxel regions).  Then, (i) to fit Gaussian profiles to the emission lines 
(with stellar fit subtracted, so they are corrected for the underlying stellar absorption, which can be a few $\rm\AA$ EW for the Balmer lines), and from this map emission-line fluxes, which show the present-day activity in the galaxy, (ii) from the model fits to the stellar continua, reconstruct a star-formation history for each spaxel. The two packages we employed in this are Porto-3D, which incorporates Starlight (Cid Fernandes et al. 2013), as used in Roche et al. (2015) and many other CALIFA survey papers, and the `Fitting Analysis using Differential evolution Optimisation' (FADO) more recently developed in Portugal (Gomes \& Papaderos 2017). 
 Both produce an output data cube where the layers are a series of 2D maps of the galaxy, depicting the flux per pixel in $\rm H\alpha$ and other lines, the spatially resolved radial velocity (from emission lines), the model-fit mean stellar age and metallicity (weighted by luminosity or mass) and other quantities.
 
 Both packages were run using the same  set of 236 single stellar populations from Bruzual \& Charlot (2003) for a Chabrier (2003) Initial Mass Function and based on Padova (1994) stellar evolution tracks. These included 59 ages (1 Myr to 13.5 Gyr) for 4 metallicities (1/50, 1/5, 1/2.5, 1.0 $\rm Z_{\odot})$. 
 Both fit the stellar continuum well and give closely similar measurements of line fluxes, kinematics etc., but might differ in the reconstructed star-formation history. FADO has the advantage of including nebular continuum in a self-consistent way, and there is evidence it is more accurate than Starlight for young and high sSFR galaxies (Cardoso, Gomes \& Papaderos 2019; Pappalardo et al. 2021; Breda et al. 2022), so it is used primarily for the results presented here.

 \section{Morphology, emission lines and ongoing star-formation}

    \begin{figure}
%   \hline
   \hskip -1.0cm
 \includegraphics[width=1.1\hsize,angle=0]{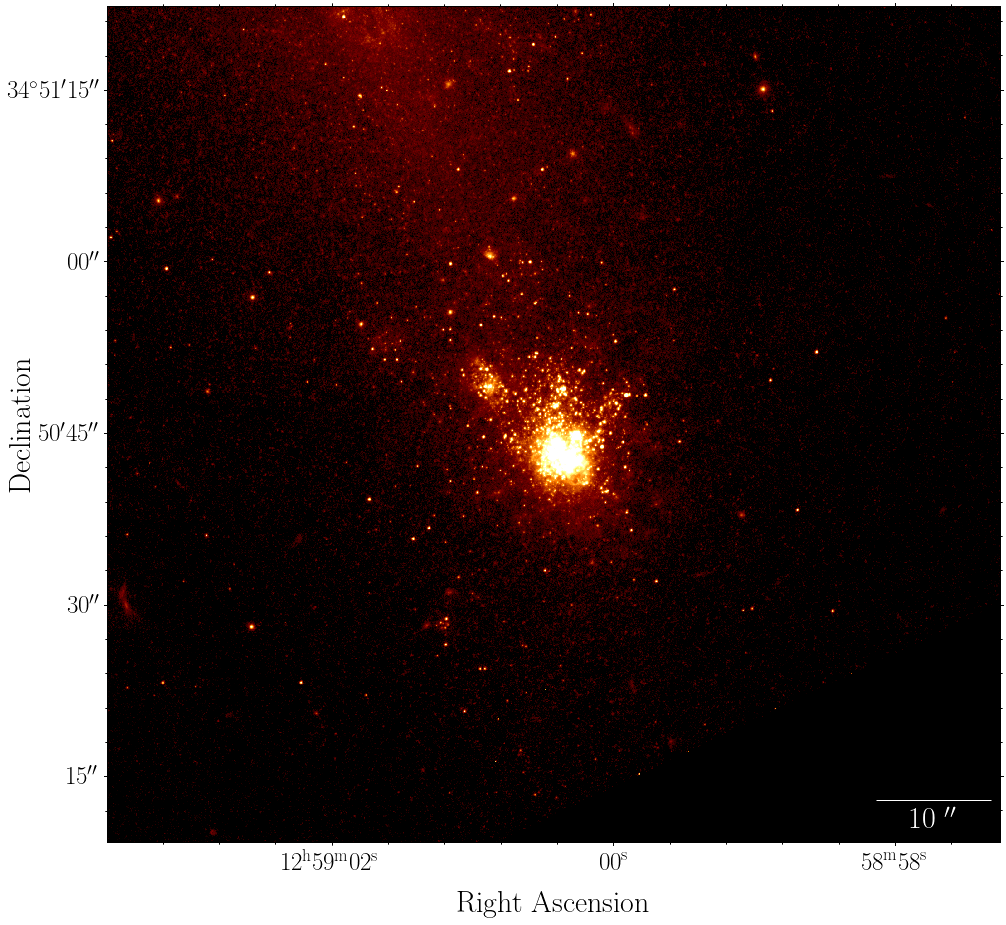}
 \vskip -0.0cm
 %\hline

\caption{HST WFC3 image centred on Mrk 59 (the southern component of NGC 4861), in broad  $I$-band (F814W), showing the $78\times 73$ arcsec area corresponding to our PMAS observation, with co-ordinates from HST astrometry. North is up, East is left} 
 \end{figure}
 
 Firstly we show (Fig 1), from  the HST WFC3 image of NGC 4861 in F814W (broad $I$-band), a cutout region $78\times73$ arcsec which corresponds to the area of our data cube. The cometary head is seen as a huge cluster, some 12 arcsec (934 pc)  in diameter, with a very high density of stars, near spheroidal but with three fainter `prongs' extending to the NE, N and NW. This is the nucleus of the southern component, Mrk 59. The centre of the NGC 4861 disk lies 1 arcmin to its NNE and is outside of our field-of-view. 
 
 Figure 2 shows the corresponding image in narrowband F658N, which includes the redshifted  $\rm H\alpha$. Although not continuum-subtracted, for this galaxy (where the $\rm H\alpha$ EW is hundreds of  $\rm\AA$ and [NII]$\lambda6584$ is much weaker) it is mostly $\rm H\alpha$ emission, from nebulae rather than directly from stars. It gives a different view of the cometary head as a spectacular giant nebula at least 1 kpc in diameter, with a complex structure containing bright  knots, and surrounded by fainter shells and filaments over half the field-of-view, tracing the entirety of the  Mrk 59 component. Further north there is a gap, and beyond this more star-forming regions in the disk of NGC 4861.

   \begin{figure}
 %  \hline
   \hskip -3.0cm

 \includegraphics[width=1.1\hsize,angle=0]{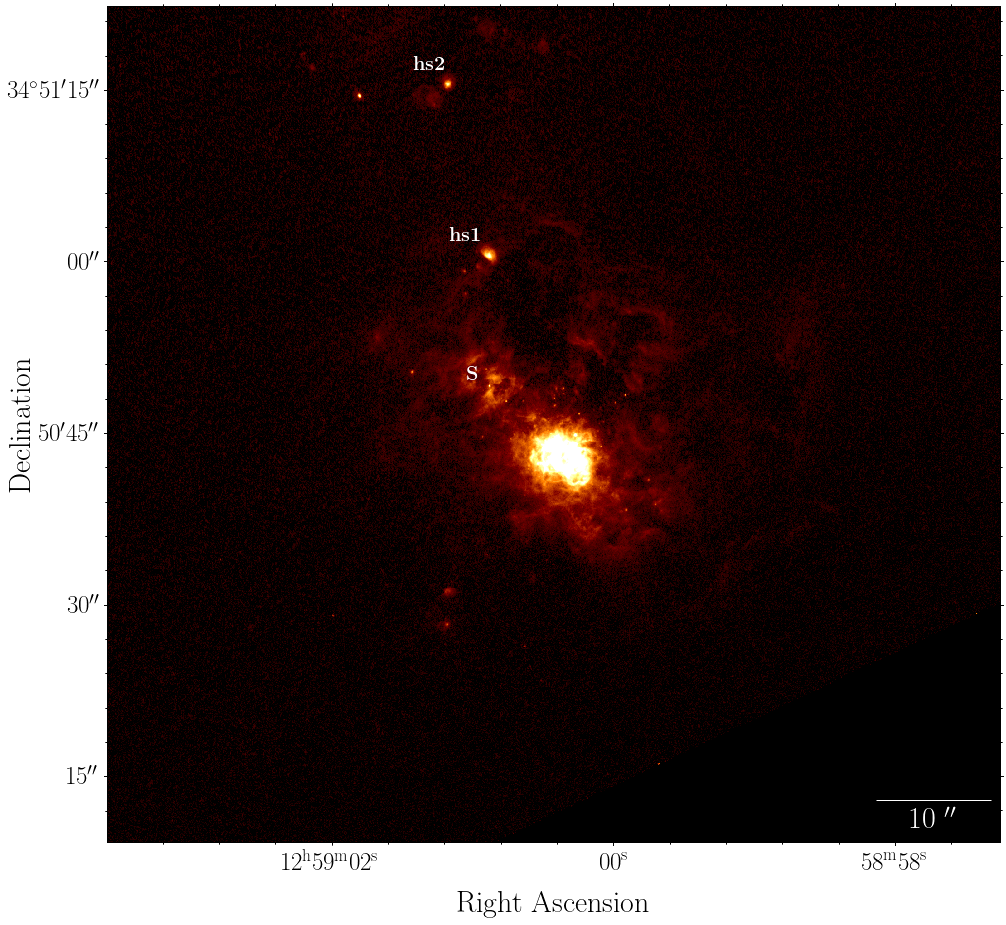}
 \vskip -0.0cm
 %\hline
 
\caption{HST WFC3 image of the same area as Fig 1 in the narrow  F658N band, primarily showing $\rm H\alpha$ emission. Labels S, hs1 and hs2 denote positions of the `spur' and `hotspot 1 and 2'.} 
 \end{figure}

 Adjacent to the central nebula there is a second region of bright extended nebulosity coinciding with the NE of the 3 stellar prongs, which can be described as a `Spur'. 
  Fig 2 shows a few other nebulae, smaller  
 ($\sim 1 $ arcsec) but high surface brightness, which can be described as `hotspots'. Imaging of a wider area (e.g. Barth et al. 1994) shows these extend in a chain of some dozen nebulae from the giant nebula across the northern part of the galaxy. The two brightest hotspots in our field of view are visible on both the HST and PMAS $\rm H\alpha$ images -- one belongs to the Mrk 59 component and lies near its northern limit, connected to the giant nebula by wispy nebulosity, the second is further north, within the  southern end of the disk component. These will be investigated individually. Hotspot 1 and 2 correspond respectively to regions 14 and 12 of Barth et al. (1994), and approximately to regions 2 and 5 on the long-slit of Noeske et al. (2000).
\begin{figure}
 \vskip -0.5cm
 \hskip -2.9cm
  \includegraphics[width=1.6\hsize,angle=0]{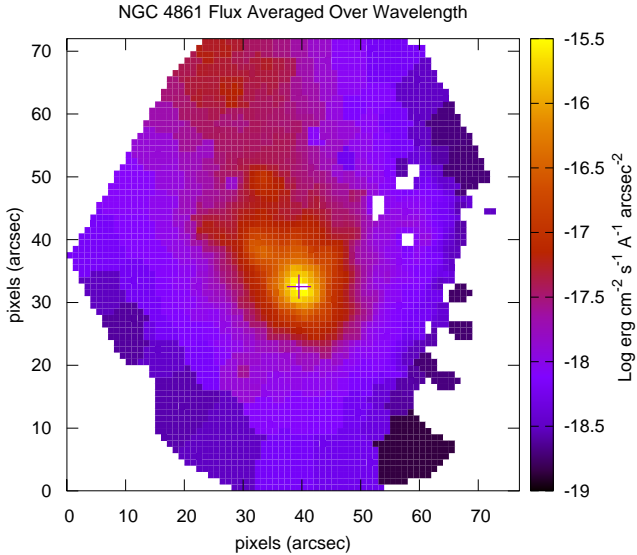}
  \vskip -0.9cm
\caption{Flux map from PMAS data, averaged over the range 3960--$\rm 6900\AA$, shown 
on log scale of flux per Angstrom.
. The cross (+) marks the central (highest flux) pixel of the Mrk 59 head/nebula, here and on our subsequent plots of IFS data, to give a common reference point.} 
 \end{figure}
  \begin{figure}
 \hskip -1cm
 \includegraphics[width=1.2\hsize,angle=0]{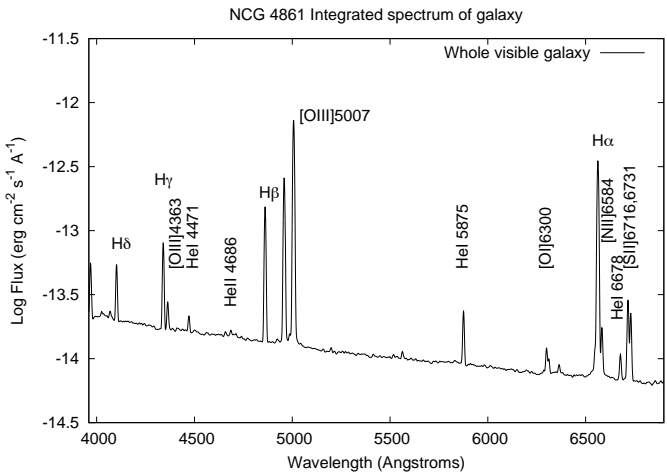}
 \vskip -0.5cm
\caption{PMAS spectrum integrated over the whole visible galaxy (the 2114 spaxels with $\rm H\alpha$ flux above a threshold $10^{-16}$ erg $\rm cm^{-2}s^{-1}$) in log scale.} 
 \end{figure}
 
 \begin{figure}
 \vskip -0.5cm
 \hskip -2.9cm
  \includegraphics[width=1.6\hsize,angle=0]{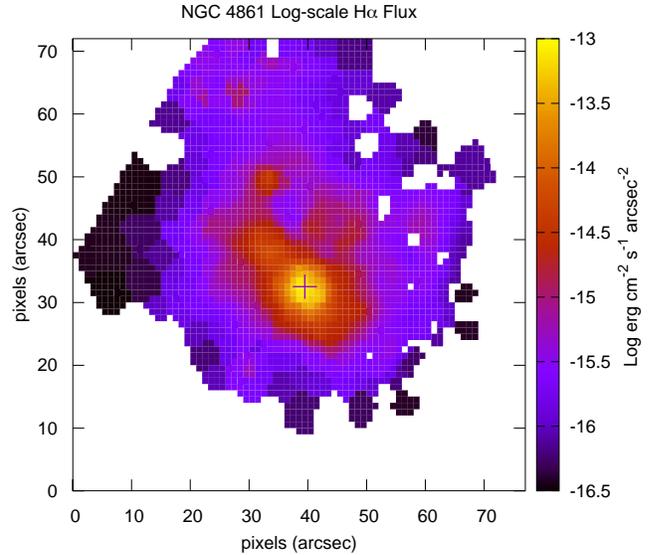}
  \vskip -0.9cm
\caption{$\rm H\alpha$ line flux map from PMAS data, on log scale.} 
 \end{figure}
Fig 3 shows our data cube averaged over the whole wavelength range $\rm 3960\AA$ to $\rm 6900\AA$. This shows the Mrk 59 nucleus, the 3 `prongs' seen in Fig 1 (faintly), and the northern component in the uppermost 15 arcsec or so. Fig 4 shows the corresponding spectrum, summed over this whole image. it shows several strong and fainter emission lines and will be discussed further in this section. 
 The $\rm H\alpha$ emission line is a good tracer of the current SFR (at least where it does not suffer a great amount of dust extinction). Fig 5 shows a map of $\rm H\alpha$ flux from our PMAS data, measured and continuum subtracted (and stellar absorption corrected) by FADO. It shows at lower resolution many of the same features as the HST image: the bright nucleus, spur and hotspots are emphasised (but not the northern component).
 
 To begin with, we define a set of regions covering the most obvious sites of ongoing star-formation, (i) an aperture of radius $r=6$ arcsec centred on the peak of continuum, located at (40,34) arcsec, covering the central bright part of the giant nebula/nucleus; an aperture of radius $r=2.5$ arcsec centred on the position (34,40) arcsec, and 2 more pixels, to cover the NE spur, and an aperture of radius $r=2$ arcsec centred on both hotspot 1and hotspot 2; Table 2 gives these positions in RA/Dec. 
   The $\rm H\alpha$ flux is summed in each of these regions (Table 2), and also for the entire galaxy within our field-of-view and down to the $\rm H\alpha$ isophote $1\times 10^{-16}$ erg $\rm cm^{-2}s^{-1} arcsec^{-2}$ (total area 2114 $\rm arcsec^2$), this is called Mrk 59 in the Table.
 These fluxes are converted to luminosity L assuming the distance modulus 31.01 mag.
 The SFR in solar masses ($\rm M_{\odot}$) per year can then be estimated as $\rm SFR=4.6\times10^{-42}L(H\alpha)$ (Cochrane et al. 2018, adapted from Kennicutt 1998 for a Chabrier 2003 Initial Mass Function ).

 \begin{table*}
\begin{tabular}{lcccccccc}
\hline
Region & R.A. & Dec & $\rm F(H\alpha)_{obs}$ & $\rm F(H\beta)_{obs}$ & $\rm L(H\alpha)^1$ & $\rm SFR^2$ & $\rm [OIII]5007/H\beta$ & $\rm [NII]6584/H\alpha$\\
  & hh:mm:ss & dd:mm:ss & \multispan{2} $10^{-16}$ erg $\rm cm^{-2}s^{-1}$ & erg $\rm s^{-1}$ & $\rm M_{\odot}yr^{-1}$ & ratio & ratio\\
    \hline
    
Nucleus $r<6$ & 12:59:00.4 & +34:50:43 & $21759(\pm 18)$ & $8996(\pm 12)$ & $6.601\times 10^{40}$ & 0.304 &  $5.80(\pm 0.01)$  & $0.024(\pm 0.001)$ \\
Spur & 12:59:00.9  & +34:50:50 & $1114(\pm 3)$ & $420(\pm 2)$ & $3.38\times 10^{39}$ & 0.0155 & $4.57(\pm 0.04)$  & $0.035(\pm 0.001)$ \\
Hotspot 1 & 12:59:00.9 & +34:51:15 & $373(\pm 3)$ & $142(\pm 2)$ & $1.13\times 10^{39}$ & 0.0052 & $4.29(\pm 0.07)$  & $0.032(\pm 0.002)$ \\
Hotspot 2 & 12:59:01.2 & +34:51:15 & $116(\pm 1)$ & $49.4(\pm 1)$  & $3.53\times 10^{38}$ & 0.0016 & $4.74(\pm 0.14)$  & $0.047(\pm 0.004)$ \\
Mrk 59 & 12:59:00.4 & +34:50:43 & $33533(\pm 25)$   & $13513(\pm 23)$ & $1.017\times 10^{41}$ & 0.468 & $5.13(\pm 0.01)$ & $0.031(\pm 0.002)$ \\
    
\hline
\end{tabular}
\caption{Fluxes (F), luminosities (L) and emission-line ratios from the regions of the galaxy described in the text (with 
RA/Dec coordinates from the HST astrometry). $^1$ $\rm L=10^{52.482}flux$; $^2$ If we instead take $\rm SFR=13.0\times10^{-42}L(H\beta)$ the SFR estimates increase a little to    0.355, 0.0166, 0.0056, 0.0019 and 0.533      $\rm M_{\odot}yr^{-1}$ } 
\end{table*}
However, while this method of SFR estimation may be valid for whole galaxies, it may give an underestimate for much smaller hotspots due to leakage of their ionizing photons into surrounding regions (e.g. Rela\~no et al. 2012).
Our summed $\rm H\alpha$ flux for the giant nebula is similar to the $2.23\times 10^{-12}$ erg $\rm cm^{-2}s^{-1}$ of Karthick et al. (2014). We find the maximum SFR density, from the pixel with the highest $\rm H\alpha$ ($8.36\times 10^{-14}$ erg $\rm cm^{-2}s^{-1}arcsec^{-2}$), attains 2.0 $\rm M_{\odot}yr^{-1}kpc^{-2}$. We compare our $\rm H\beta$ flux with previous observations by summing over smaller areas of central nebula matching the relevant spectroscopic slits, and come close to the $3.38\times 10^{-13} $ erg $\rm cm^{-2}s^{-1}$ of Karthick et al. (2014), while our flux is double that given by Noeske et al. (2000).

The ratio $\rm H\alpha/H\beta$ (Balmer decrement) is commonly used as an estimator of internal dust extinction, which reddens the spectrum and thus increases this ratio above the theoretically expected 2.79--2.86 (Osterbrock 1989). However, our data gave Balmer decrements below this range for the majority of spaxels (e.g. the mean for the spaxels in the nucleus region was 2.54). We checked our flux calibration by examining the ratios of 5 HeI lines (4471, 4922, 5976, 6678, $7065 \rm\AA$), and found these were consistent with theoretical expectations, as was $\rm H\gamma/H\beta$, and as noted above our integrated $\rm H\alpha$ and $\rm H\beta$ fluxes look consistent with previous observations. Sub-theoretical $\rm H\alpha/H\beta$ ratios  have been reported in a number of low-metallicity starburst galaxies (e.g. Guseva et al. 2003, Gao et al. 2017)  
and have not been fully explained. Other studies of this and similar galaxies have found very little internal dust extinction (e.g. Karthick et al. 2014). Without additional observations, we can only (following other authors) use the observed fluxes and assume no significant internal dust extinction.
A few (7) central nebula spaxels with the highest $\rm H\alpha$ fluxes had particularly low Balmer decrements, as low as 2. In view of this and the unusual ratio in general, we give additional (slightly higher) SFR estimates based instead on our observed $\rm H\beta$ (Table 1), with $\rm H\alpha/H\beta$ fixed at a theoretical 2.82 for the $\rm SFR-H\alpha$ relation of Cochrane et al. (2018), so that $\rm SFR=13.0\times10^{-42}L(H\beta)$. 

\begin{figure}
\hskip -0.5cm
  \includegraphics[width=1.1\hsize,angle=0]{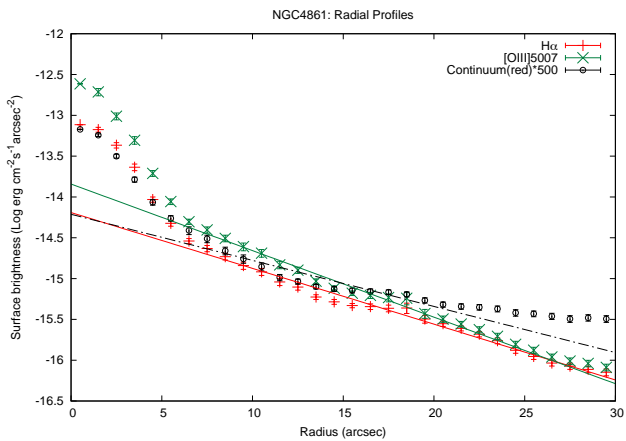}
 \vskip -0.5cm
 
\caption{Radial profile (measured in $\Delta(r)=1$ arcsec circular annuli) of the $\rm H\alpha$ (red) and
$\rm [OIII]\lambda5007$ (green crosses) emission lines, and the continuum at 6390--$\rm 6490\AA$ (black), measured in flux $\rm \AA^{-1}$ and plotted multiplied by 500 for ease of comparison with the strong lines). Our fitted exponentials are also plotted as the straight red ($\rm H\alpha$), green ([OIII]) and black (cont.) lines}.
 \end{figure}
Figure 6 shows a radial profile centred on the nucleus and calculated in circular apertures, for the red continuum (6390--6490$\rm\AA$) and the two strongest lines $\rm H\alpha$ and [OIII]$\lambda$5007. All 3 show approximately exponential profiles with a large central excess and steepening at $r<6$ arcsec, produced by the (1 kpc diameter) central nebula (as in the profiles shown by Noeske et al. 2000). At 6 to 22 arcsec we fit exponential scale lengths $r_{exp}\simeq 6.51$, 5.38 and 4.51 arcsec for continuum, $\rm H\alpha$ and [OIII]$\lambda5007$, respectively. The half-light radii are 8.5, 4  and 3 arcsec. Mrk 59, whether disk-like or more irregular, is small with $r_{exp}\simeq 0.5$ kpc. The $\rm [OIII]\lambda 5007$ emission is more centrally concentrated than the Balmer lines, and this leads to consideration of the excitation ratio, $\rm [OIII]\lambda5007/H\beta$.

We calculate this ratio for each spaxel and region by summing the [OIII]$\lambda5007$ and $\rm H\beta$ emission fluxes (not including [OIII]$\lambda4959$) from the FADO fits (stellar components subtracted), to map the excitation for the whole galaxy (Table 2, Fig 7). 

    \begin{figure}
    \hskip -2.9cm
 \includegraphics[width=1.6\hsize,angle=0]{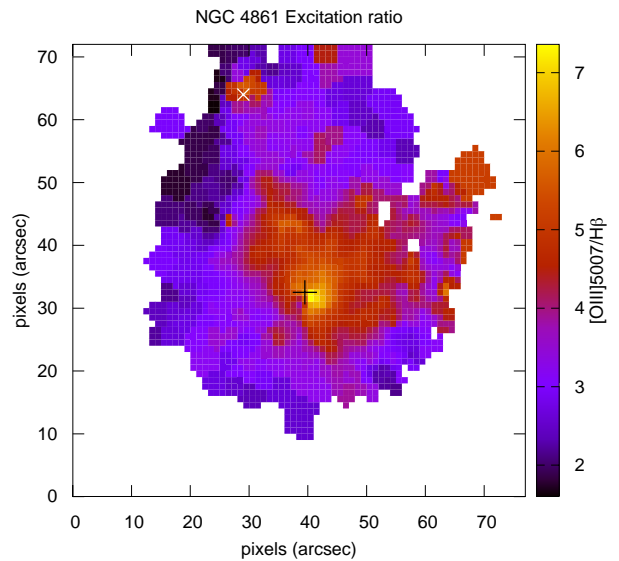}
 \vskip -1.2cm
\caption{Map of the excitation ratio $\rm [OIII]\lambda5007/H\beta$, showing highest values ($\sim 7$) at the giant nebula and a distinct second peak in the north at a position corresponding to Hotspot 2 (small white X).} 
 \end{figure}

There is a high peak in excitation ratio at the giant nebula, reaching a maximum of (at about an arcsec SW of the peak of flux) of 7.35, which is in the range of `green pea' galaxies. The ratio decreases with distance from the nucleus, remaining $>4.3$ over  most of the Mrk 59 component, extending to  Hotspot 1. In faint outer regions  
($\rm F(H\alpha)\leq 10^{-15}$ erg $\rm cm^{-2}s^{-1}arcsec^{-2}$) it is lower, between 3 and 4, and decreases further into the northern component, except for an obvious second peak at the position of Hotspot 2, where excitation reaches a maximum of 5.64. This was noted by Barth et al. (1994), who found their corresponding region 12 lay well above the radial gradient in excitation traced by the other emitting regions, and they suggested this could be attributed to shocks produced by a supernova. Hotspot 2 looks elongated E-W on the excitation map, and the HST image (Fig 2)  shows two structures here, spanning 3 arcsec, the western sharply peaked and the eastern more diffuse and shell-like. 

    \begin{figure}
    \hskip -2.9cm
 \includegraphics[width=1.6\hsize,angle=0]{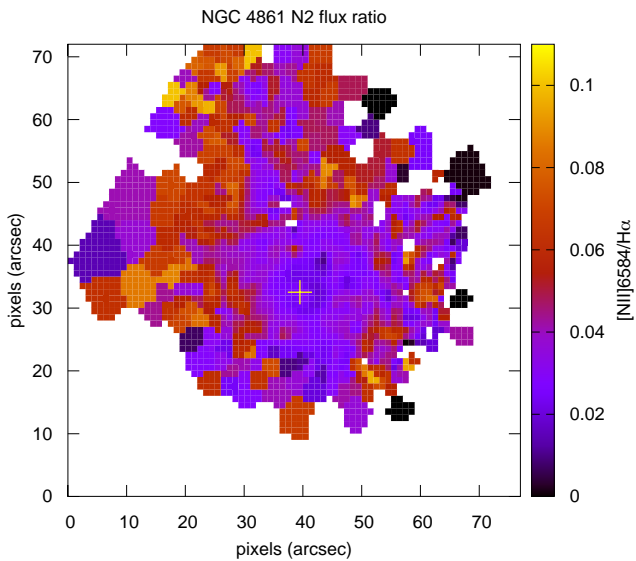}
 \vskip -1.2cm
\caption{Map of the metallicity-sensitive N2 flux ratio $\rm [NII]6584/H\alpha$.}
 \end{figure}

Fig 8 maps another key line ratio, $\rm [NII]\lambda6584/H\alpha$ (N2), which, in contrast to the excitation, shows no central peak at all. This is evidence against AGN emission in Mrk 59.  The N2 map, noisy because the [NII] lines are weak, shows no strong features but looks slightly higher in the far north. 
By the N2 and O3N2 (dust insensitive) metallicity estimators as calibrated by Marino et al. (2013), the line ratios point to $\rm 12+log(O/H)=7.99$ (N2) or 8.02 (O3N2) for the giant nebula (in agreement with Noeske et al. 2000) and 8.05/8.06 for the `whole' galaxy. Note that at this low metallicity we are  outside the fitted range for the O3N2 estimator, while N2 continues to follow a linear relation (Marino et al., Fig 4), so may be preferable.

  \begin{figure}
     \hskip -2.9cm
 \includegraphics[width=1.6\hsize,angle=0]{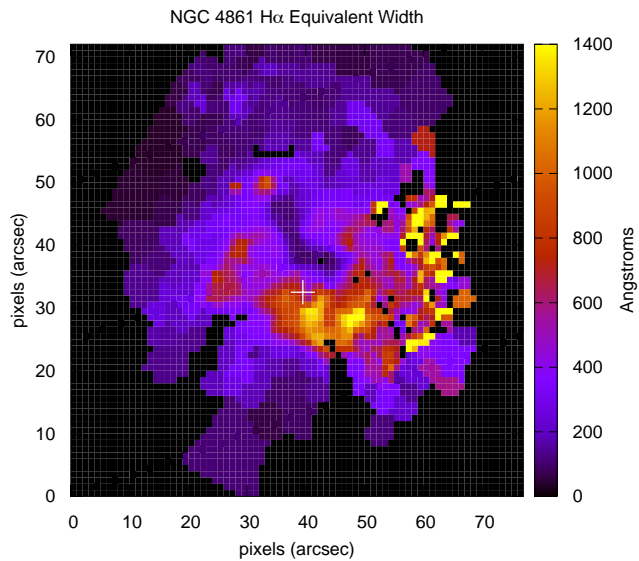}
\vskip -1.2cm
\caption{Map of $\rm H\alpha$ emission equivalent width.}
 \end{figure}
 Fig 9 shows a map of the $\rm H\alpha$ EW. It is
 very high compared to most galaxies, $\geq 400\rm \AA$ over Mrk 59, and at the nucleus it approaches $700\rm\AA$. However, it shows a different pattern from the centrally-peaked line fluxes, in that the highest EWs of all (800--$1400\rm \AA$) are found to the south and west of the central point, at the edge of the central nebula, and extend into lower surface brightness nebulosity further west. The EW is sensitive to stellar age, e.g. according to Pappalardo at al. (2021) $\rm 1000\AA$ corresponds to 5.6 Myr  of a constant SFR, so this is another region with very young stars.
 
  \begin{figure}
 \hskip -1cm
 \includegraphics[width=1.1\hsize,angle=0]{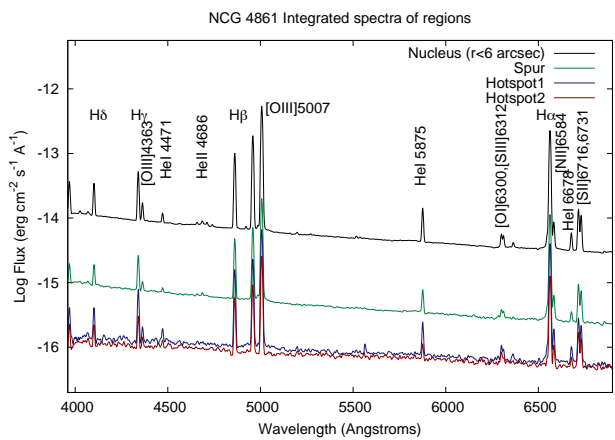}
 
\caption{PMAS spectra integrated over the nucleus, spur and hotspot star-forming regions, flux in log scale.} 
 \end{figure}

Fig 10 shows the spectra of the nucleus, spur and hotspot regions.  All have blue continua with strong emission lines and a lack of absorption features. Auroral lines [OIII]$\lambda4363$ and 
[SIII]$\lambda6312$ are visible, indicators of high electron temperatures. The two star-formation Hotspots have similar continua but the Hotspot 1 has stronger emission lines (and is slightly bluer). Of particular interest in the nucleus spectrum is a close group of 4 small emission lines centred on HeII$\lambda4686$ -- this is the distinctive WR star signature known as the `blue bump' and will be investigated in section 6. 

 \section{Kinematics}
   
       Fig 11 shows a map of gas radial velocities, measured from (FADO) fits to the strongest emission lines, given relative to the systematic (i.e. centre of the disk) recession velocity 790 km $\rm s^{-1}$.
        On this scale the velocities in our field-of-view are all negative with the centre of the nebula at  about $-37$ km $\rm s^{-1}$. Uncertainties are $\sim10$ km $\rm s^{-1}$. 
   High velocity ($>100$ km $\rm s^{-1}$) outflows are not seen anywhere.
   The strongest feature is a blueshifted (by 40--50 km $\rm s^{-1}$) region $\sim 12$ arcsec  NW of the nucleus. Further NW and almost to the image edge is a $\sim 20$ km $\rm s^{-1}$ redshifted region. These appear to be motions in the wispy filaments faintly visible in the HST $\rm H\alpha$ image (Fig 2). In our PMAS $\rm H\alpha$ image (Fig 5) we can also see faint emission here, forming a sort of loop.  These features appear to correspond spatially to the `SGS4 shell' described by van Eymeren et al. (2007 and 2009) -- in $\rm H\alpha$ and HI they found motions of $\simeq 30$ km $\rm s^{-1}$ in this region W of the galaxy. Presumably this is an expanding shell of gas driven by the central starburst, now at least 1 kpc in size but very faint, with $\rm H\alpha$ surface brightness $\sim 5\times10^{-16}$ erg $\rm cm^{-2}s^{-1}arcsec^{-2}$.  There is also  some sign of a blueshifted region  S/SE of the nucleus, where the HST image shows 2 small emitting nebulae. 
    \begin{figure}
    \vskip -1.0cm
     \hskip -2.9cm
 \includegraphics[width=1.6\hsize,angle=0]{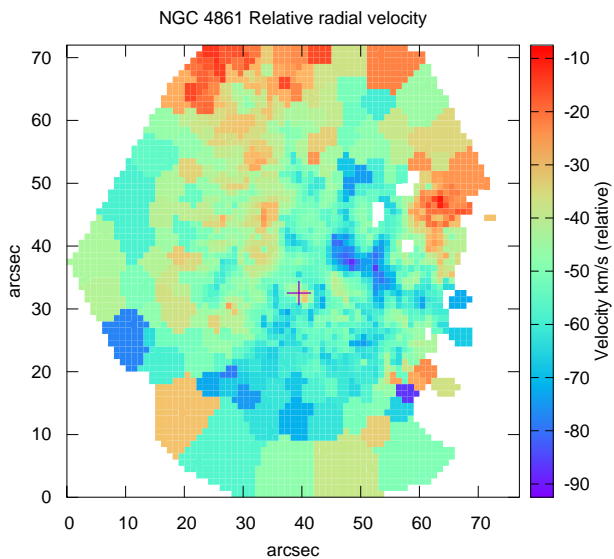}
 \vskip -1.2cm
\caption{Map of Radial velocity in km $\rm s^{-1}$, relative to galaxy systematic velocity, measured from the emission lines in each spaxel (by FADO).} 
 \end{figure}
 
  A large region in the far north of the field-of-view is redshifted by $\sim 25$ km $\rm s^{-1}$ relative to the southern component.  This shift is the abrupt velocity difference between the rotating northern disk and Mrk 59, seen by  Thuan, Hibbard and L\'evrier (2004). Hotspot 2 looks 10--15 km $\rm s^{-1}$ less redshifted than surrounding pixels, perhaps this is evidence for a small outflow but higher spectral resolution is needed to confirm this.
   
 In summary,  we see in this galaxy the motion of large faint shell structures, at velocities too low to drive gas out of the system ($v_{escape}\simeq 160$ km $\rm s^{-1}$; van Eymeren et al. 2007), and near the north edge, the disk rotation. There is no sign of a centrally concentrated wind or high velocity outflow like those seen in some merging galaxies (e.g. Wild et al. 2014, Roche et al. 2015), which seem to be in a later stage of evolution.
   
  We use the kinematic map to define another two regions for spectral analysis. The first is the redshifted (relative velocity $>-26$ km $\rm s^{-1}$) region in the north/NE of the image, which is essentially the part of the northern disk galaxy  (totalling 115 $\rm arcsec^2$) within our 
  field-of-view (but excluding hotspot 2 because this is less redshifted, and is being considered separately). The second is the most blueshifted ($-90<\Delta(v)<-70$ km $\rm s^{-1}$) part of the low-surface-brightness region west of the giant nebula (52 $\rm arcsec^2$).     Also we define a seventh region, of the 40 spaxels with the highest $\rm EW(H\alpha)$ of $\rm >1000\AA$ and which lie outside of the $r<6$ arcsec nucleus region, but extend to its south and west (Fig 9).
  Table 3 lists some properties of these additional regions. Fig 12 shows the positions of all regions.
   \begin{table*}
\begin{tabular}{lcccccc}
\hline
Region & $\rm F(H\alpha)_{obs}$ & $\rm F(H\beta)_{obs}$ & $\rm L(H\alpha)$ & SFR & $\rm [OIII]5007/H\beta$ & $\rm [NII]6584/H\alpha$\\
  & \multispan{2} $10^{-16}$ erg $\rm cm^{-2}s^{-1}$ & erg $\rm s^{-1}$ & $\rm M_{\odot}yr^{-1}$ & ratio & ratio\\
    \hline
    
Redshifted & $348(\pm 3)$ & $135(\pm 3)$ & $1.06\times 10^{39}$ & 0.0049 &  $1.97(\pm 0.06)$  & $0.068(\pm 0.005)$ \\
Blueshifted & $381(\pm 3)$ & $147(\pm 3)$ & $1.16\times 10^{39}$ & 0.0053 &  $4.66(\pm 0.11)$ &  $0.034(\pm 0.004)$ \\
$\rm EW>1000\AA$ & $954(\pm 5)$ & $363(\pm 3)$ & $2.89\times 10^{39}$ &  0.0133 &
 $4.54(\pm 0.05)$ & $0.033(\pm 0.001)$ \\
\hline

\end{tabular}
\caption{Line fluxes and ratios of an additional 3 regions. If we estimate SFR from $\rm H\beta$ rather than $\rm H\alpha$, using again  $\rm SFR=13.0\times10^{-42}L(H\beta)$, the SFR estimates increase a little to    0.0053, 0.0058 and 0.0143 $\rm M_{\odot}yr^{-1}$.}
\end{table*}

The blueshifted and high-EW regions have a high excitation and low N2 ratio, like all other parts of the Mrk 59 structure. The blueshifted region spectrum (Fig 13) is similar to that of hotspot 1, another outlying part of Mrk 59. It seems to be just a diffuse part of Mrk 59 which, due to the expanding shells around the nebula, happens to have acquired the highest relative radial velocity. The high-EW region has a similar spectrum to the blueshifted region, which it is close to, except all the emission lines (not only $\rm H\alpha$) are about twice as strong relative to the continuum. 

\begin{figure}
\vskip -1.0cm
\hskip -2.9cm
 \includegraphics[width=1.6\hsize,angle=0]{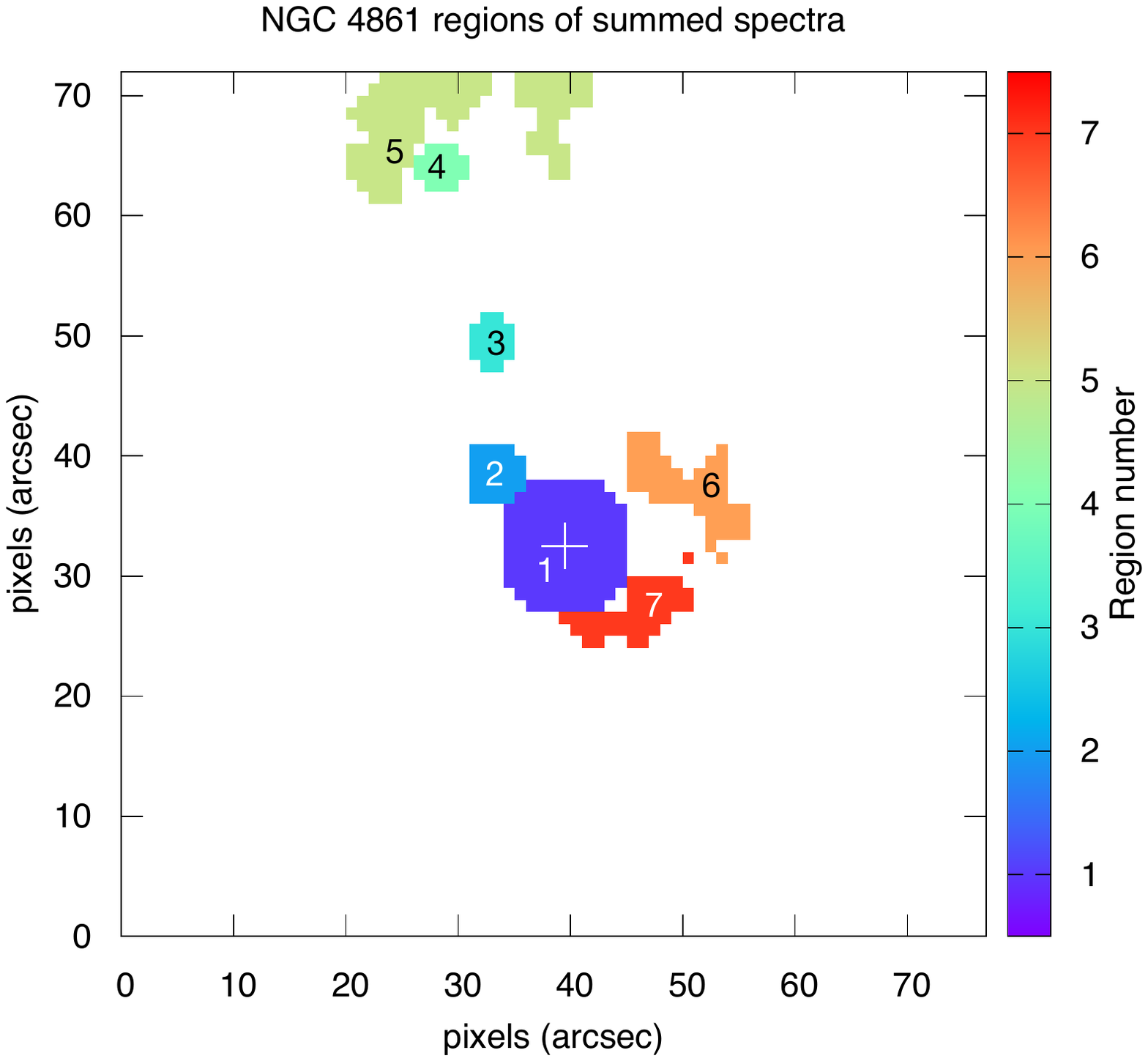}
\vskip -0.9cm
\caption{Positions of the 7 regions described above: 1 being the central giant nebula, 2 the spur, 3 is hotspot 1, 4 is hotspot 2, number 5  a redshifted region from the old part of the galaxy, 6 a blueshifted diffuse region, 7 a region of spaxels with the highest $\rm H\alpha$  EWs, close to the nucleus.}
 \end{figure}

       \begin{figure}
 \hskip -1cm
 \includegraphics[width=1.1\hsize,angle=0]{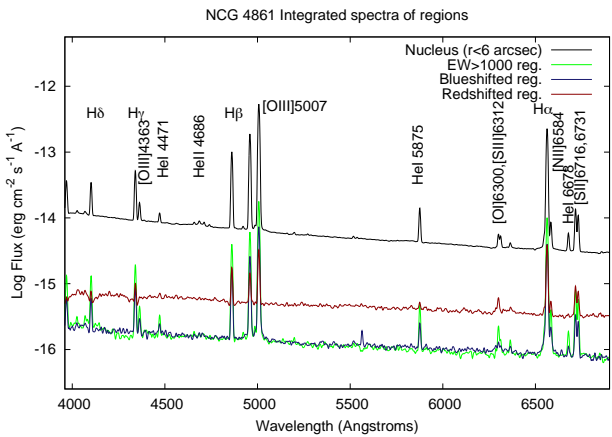}
 
\caption{PMAS spectra integrated over the blueshifted and redshifted regions described, together with the nucleus and high-EW region, flux in log scale.} 
 \end{figure}

The redshifted region, part of the northern NGC 4861 disk, has a lower excitation than anywhere in Mrk 59 but a higher N2 ratio, suggesting a higher metallicity -- the Marino et al. (2013) relations estimate  $\rm 12+log(O/H)=$ 8.20 (by N2) and 8.22 (by O3N2);  consistent with Noeske et al. (2000, Fig 12) at this position. It has weaker emission lines than the blueshifted region but is certainly not `red and dead', as the $\rm H\alpha$ EW is still $\sim 100\rm\AA$.  The redshifted region also shows stronger $\rm
 [OI]\lambda6300$ emission relative to the other lines ($\rm EW\simeq 8\AA$, flux $\rm 8\%$ of $\rm H\alpha$) than elsewhere in this galaxy, and deep Balmer absorption features ($\rm H\delta_{abs}\simeq 5.5
$--$10\rm\AA$)  coinciding with narrower emission lines (in $\rm H\beta,\gamma,\delta$).
[OI]$\lambda$6300 might indicate shocks, and the Balmer absorption lines a post-starburst region. Hotspot 2 which lies adjacent has more than twice the excitation. These features will be compared further in the next section.

 \section{Spatially Resolved Star-formation History}
  Our FADO model estimates the total stellar mass present within our field-of-view as $1.38\times 10^8$ $\rm M_{\odot}$. Dividing the $\rm H\alpha$ or $\rm H\beta$ based SFRs by this mass gives 
the specific SFR 3.4--3.9 $\rm Gyr^{-1}$ (Porto 3D estimated a slightly greater stellar mass, perhaps because of its non-inclusion of nebular continuum in the fit -- $1.67\times 10^8$ $\rm M_{\odot}$).
Each spaxel typically represents of order $\rm \sim 10^{4.6-5}~M_{\odot}$. This specific SFR means
Mrk 59 is undergoing a starburst, as even if it is only 1 Gyr old its current SFR is well above its time-averaged SFR. These properties are within the range of `green pea' galaxies (Izotov et al. 2011), however the relatively low stellar mass and SFR of Mrk 59 place it near the upper limit of the `blueberry' class of smaller starburst galaxies (Liu et al. 2022, Paswan et al. 2022).

We run Porto3D and FADO to fit individual spectral synthesis models (and therefore star-formation histories) to the observed spectra in each of the 1312 single-spaxel  or larger Voronoi elements. This maximizes the spatial sampling but at the expense of low signal/noise spectra and hence noisy stellar age estimates. However, this may be sufficient to make an approximate age map for the whole galaxy. To begin with, each of the fitted star-formation histories are characterized by two numbers, a luminosity-weighted mean age ($t_L$) with $L$ calculated at $\rm 5080\AA$ and a mass-weighted mean age ($t_M$). 
We show here the mean age maps produced by FADO (Figures 14 and 15). They are noisy but do show a common 2D pattern.
 
   \begin{figure}
  \vskip -1.0cm
    \hskip -2.9cm
 \includegraphics[width=1.6\hsize,angle=0]{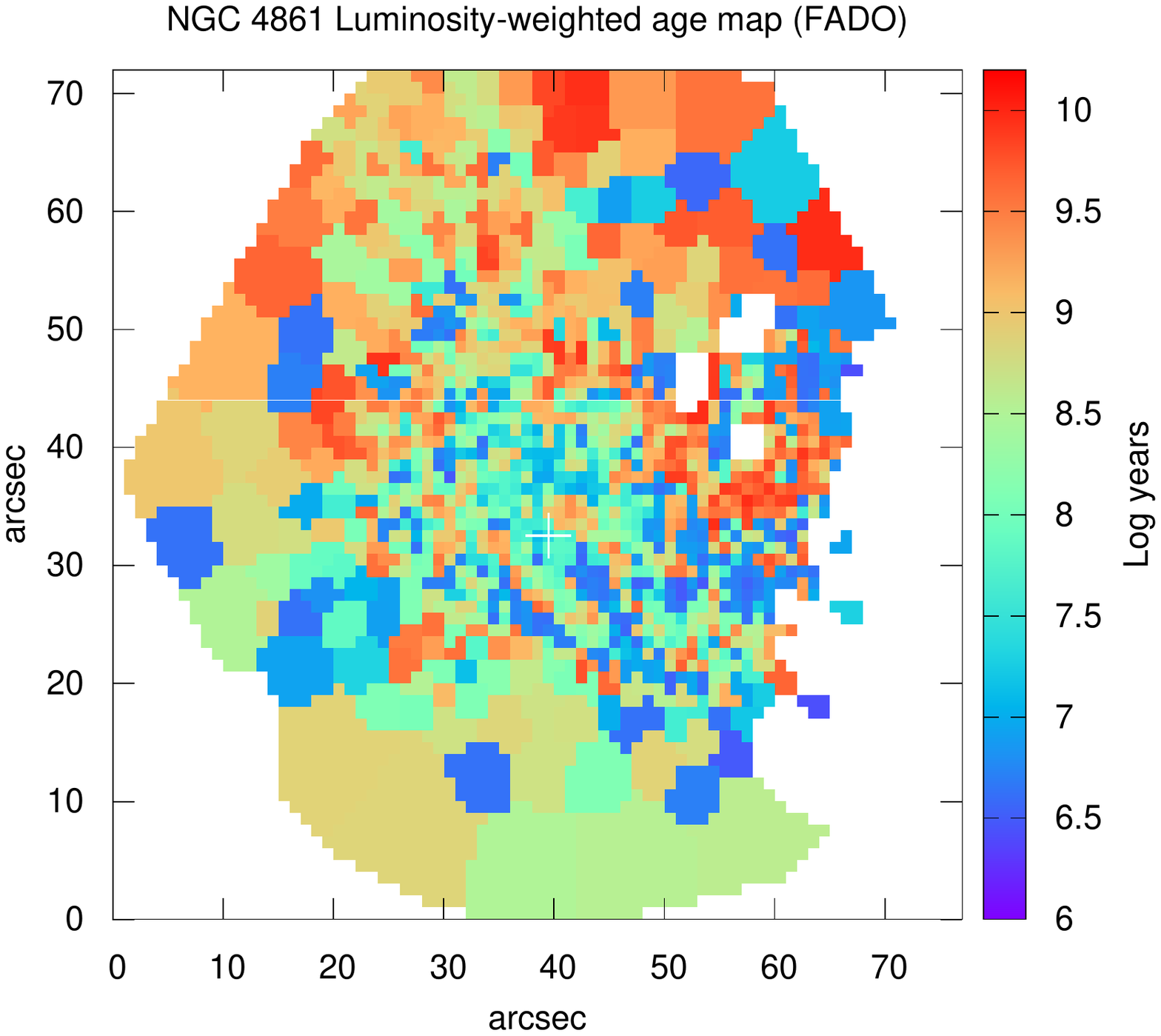}
 \vskip -1.3cm
\caption{Luminosity-weighted stellar age in each spaxel/Voronoi element (on log scale), as estimated using FADO.} 
%\end{figure}

%\begin{figure}
\hskip -2.9cm
 \includegraphics[width=1.6\hsize,angle=0]{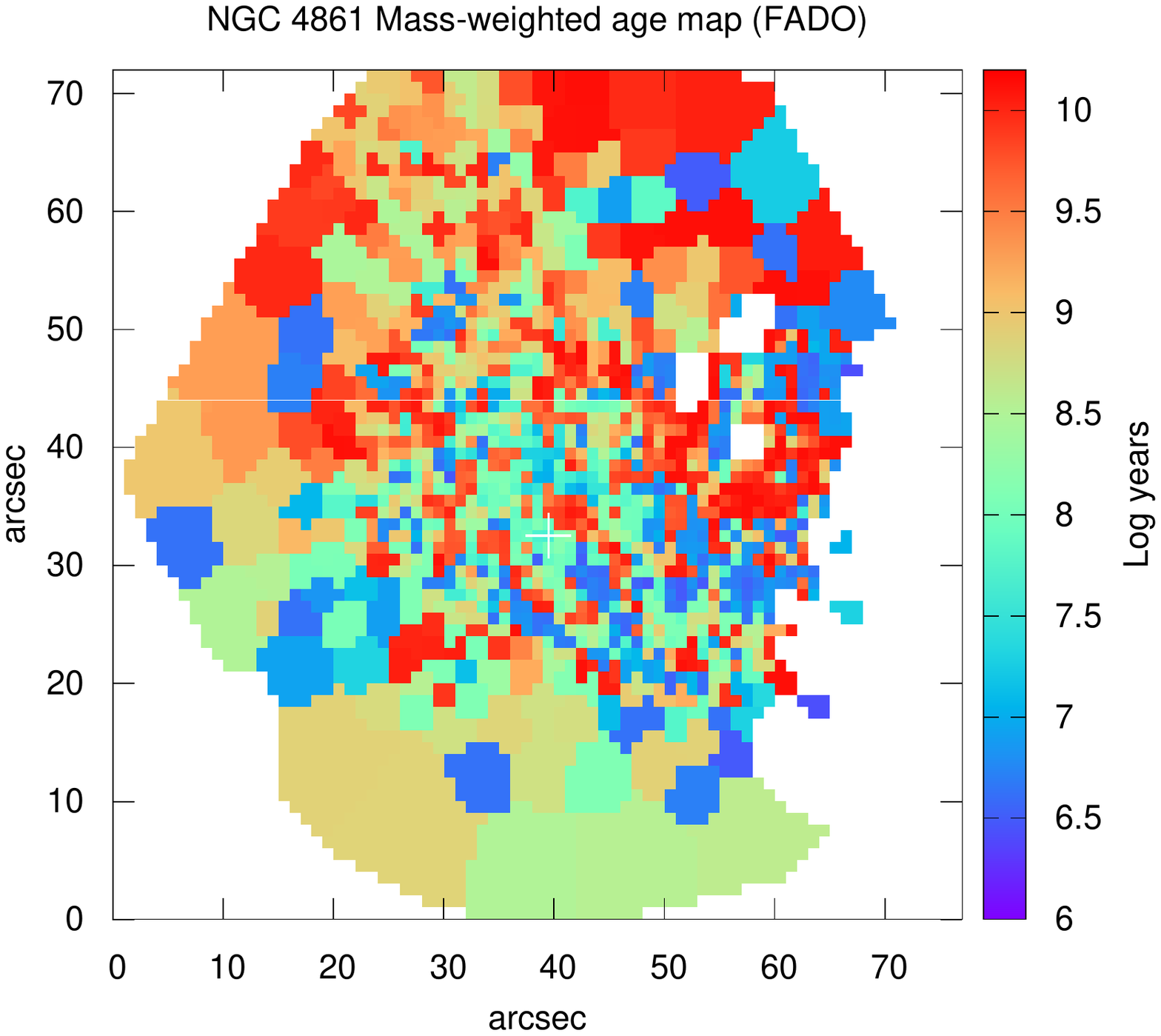}
 \vskip -1.3cm
\caption{Mass-weighted stellar age in each spaxel/Voronoi element (on log scale), as estimated using FADO.} 
 \end{figure}

The $t_M$ map tends to show greater ages than $t_L$, and even more noise, as expected if the star-formation histories are quite extended. Table 4, in columns 3 and 4, gives the mean log stellar ages (and errors on the means from pixel-to-pixel scatter) from the two maps shown, for each of the 7 regions  (Fig 12). By averaging over regions it can be seen that both maps show a $\sim 1$ dex difference between the giant nebula in the 
lower-centre  ($\sim 10^{8}$yr)  and the older north and NE, i.e. the northern disk component ($\sim 10^9$ yr). 
In averaged ages, the nucleus, spur and high-EW region are youngest, the hotspot 1 and blueshifted region older, the northern hotspot 2 and the redshifted region older still.

      FADO was then run on the summed spectra of each of the 7 regions -- which will have a much higher signal/noise than a single-spaxel spectrum and should therefore give better age-resolution. For each region spectrum, FADO estimated a star-formation history shown in a pair of plots. The upper is the luminosity fraction of each fitted stellar component, and the lower the mass fraction on a log scale, both against age (the curves are smoothed versions of the population vectors). The first six are shown on Fig 16.

Firstly, the nucleus (region 1) and spur (2) are found in our analysis to have similar histories, formed in continuous bursts over the past $10^8$ years, in fact slightly more, with both showing large components at age 125 Myr.  Both have a substantial  content of 
 $<10\rm~ Myr$ age stars, estimated as 8\% by mass in region 1 and about 20\% in region 2, which implies the SFRs are not decreasing. No older stars at age $>0.3$ Gyr are detected.

  The hotspot 1 is seen to have a different, bimodal history, with some very young stars ($\rm <10~Myr$) as expected from the strong emission lines, and an underlying older component with age 1 Gyr, which might rather be called intermediate age.  The young stars make up only 
 about 3\% of the mass but 55\% of the luminosity.

 Hotspot 2 is similarly bimodal with a smaller fraction of very young stars (1\% by mass, 30\% luminosity) and the median ($t_{1/2}$) age of the older component at 1.75 Gyr; the starburst is sufficient to locally double the  $\rm [OIII]/H\beta$ ratio.
 The redshifted region (5) -- representing the northern disk component and not a concentration of star-formation like hotspot 2 --  is made up of the intermediate age stellar population, midpoint age 1.15 Gyr, with only a tiny amount (2\% by luminosity) of young stars. However, we had found a greater $\rm H\alpha$ flux, and therefore estimated a higher SFR, for region 5 than for hotspot 2 (Table 2 and 3). It is possible Hotspot 2 has a higher SFR than our estimate but is leaking ionizing photons a few hundred pc into region 5, causing displaced $\rm H\alpha$ emission. From Rela\~no et al. (2012) this leakage could be $\sim25\%$ for compact regions. On the other hand, region 5 contains at least one small emission nebula of its own.

      The blueshifted region (6) shows an odd trimodal age distribution in our fit, with peaks at 3 Myr, 100 Myr and 6 Gyr.  There is again a high luminosity fraction (29\%) of very young stars, but the fit 
 differs from the other regions in that it includes a large fraction of stars older than 3 Gyr. FADO fits to other `green pea' galaxies have in a few cases shown a component of very old ($\rm \sim 10~Gyr$) stars (Fernandez et al. 2022), but it also seems possible the fit for the older stars is less accurate here, perhaps because of the low surface brightness, and the age is overestimated.

 For the region (7), selected as having the highest $\rm H\alpha$ EWs, the derived star-formation history (Fig 17) is very similar to hotspot 1, with a mixture of very young ($<10^7$ yr) and intermediate age  (0.5--1.0 Gyr) stars, the first being about 5\% by mass and   24\% by luminosity. The high EW does not necessarily mean this part of the galaxy is very young but rather reflects a high content of very young stars from the current starburst.
 
 Finally, we analyse the `whole galaxy' spectrum (plotted in Fig 4) made up by summing all 2114 spaxels with $\rm H\alpha$ flux $\geq 1\times 10^{-16}$ erg $\rm s^{-1} arcsec^{-2}$ (same as the `Mrk 59' area of Table 2).
 This could represent how a small galaxy might look if placed at high redshift where only a single spectrum could be obtained, rather than an IFS cube. About half of the flux in this spectrum comes from the nucleus and adding in the remainder of the galaxy makes it redder. Its FADO fitted age distribution resembles that of the nucleus alone, with a higher ratio of $10^8$ yr age relative to very young stars. 
 It has no $>0.3$ Gyr old stars, despite that a $\sim 1$ Gyr age component was fitted as the most massive for 4 of our region spectra. This shows how attempts to reconstruct star formation histories for high redshift galaxies may be biased and represent mostly the bright nuclei -- but also hints that more complex pictures might emerge when it is possible to employ higher-resolution IFS (e.g. with the JWST). 
 
 It is not simple to estimate accurate uncertainties on the stellar age estimates or star-formation histories from this type of complex analysis. FADO has been claimed to have a formal  uncertainty on the  age of only $\Delta \rm log( t_L ) \sim 0.005$ or 1\% (Fernandez et al. 2022), however the simulations of Pappalardo et al. (2021)  suggested  $\Delta \rm log( t _L) \sim 0.03$ or 7\% for high signal-to-noise ratio spectra. Uncertainties will depend on quality of spectra and other factors  that vary between galaxies or stellar populations. This is outside the scope of this paper and is being investigated by Papaderos et al. (in prep).  FADO appears to work well for extreme emission line galaxies (Breda et al. 2022). It is encouraging we obtain a fairly consistent picture of the spatially resolved star formation history from analysis of multiple and separated regions.
 
 In general we see  an underlying $\sim 1$ Gyr age (intermediate) population (range 0.5--2.0 Gyr), which could be the NGC 4861 disk (centred northwards of our field-of-view), plus a different distribution of very young stars (Mrk 59)  centred instead on the giant nebula and with a  luminosity fraction varying from 2\% (region 5) to near 100\%. This age bimodality resembles that of Noeske et al. (2000, Table 6) who fitted their long-slit spectrum/colours of this galaxy with combinations of 4--25  Myr and 1--2 Gyr age stars, the former dominating in the nucleus and their region 2 (our hotspot 1) and the latter northwards.

\begin{table*}
\begin{tabular}{lcccccccc}
\hline
Region & $N_{pix}$ & mean log $t_L$ & mean log $t_M$ & region $t_L$ & region $t_M$ & $\rm H\alpha$ EW & $\rm 12+log10(O/H)$ & log $\rm M_{*}$\\
   & $\rm arcsec^2$ & log yr & log yr & Myr & Myr & $\rm \AA$  & by N2 & $\rm M_{\odot}$ \\
  \hline
  Nucleus & 109 & $7.92\pm0.08$ & $8.31\pm 0.10$ & 75 & 105 & $664(\pm 3)$  & 7.99  & 7.14 \\
  Spur & 23 & $8.11\pm 0.16$  & $8.69\pm0.22$ & 39 & 90 & $415(\pm 2)$ & 8.07 & 5.99 \\
   Hotspot 1 &  16 & $8.58\pm 0.17$ &  $8.89\pm 0.20$ & 404 & 879 & $617(\pm7)$  & 8.05 & 5.91\\
   Hotspot 2 &  16  & $8.94\pm 0.15$ &  $9.25\pm 0.19$ & 975 & 1666 &  $213(\pm 3)$ & 8.13 & 5.99\\
   Redshifted region & 115 & $9.13\pm 0.04$ & $9.36\pm 0.04$ & 1140 & 1813 & $103(\pm 1)$ & 8.20 & 6.90\\
   Blueshifted region & 52 & $8.52\pm 0.16$ & $8.80\pm 0.18$ & $2619^1$ & $6011^1$ & $435(\pm 5)$ & 8.06 & 6.50 \\
   Highest EW region & 40 & $7.93\pm0.16$ & $8.24\pm 0.20$ & 337 & 446 &$ 1191(\pm 5)$ & 8.06  & 5.95\\
   \hline
      \end{tabular}
      \caption{(i) Region; (ii) size of each region in pixels or $\rm arcsec^2$; (iii) mean log $t_L$ and (iv) mean log $t_M$ (with error of mean) for each region from averaging over the region in the FADO age maps, Figs 14 and 15; (v) Stellar ages $T_L$ and (vi) $t_M$ from FADO fits to the whole-region integrated spectra (Fig 16 and 17); (vii) $\rm H\alpha$ EWs from FADO fits to whole region spectra; (viii) metallicity estimated from strong line fluxes (N2: from line ratios in Table 2 and 3), statistical uncertainties are about $\pm0.01$, but note the Marino et al. 2013 fit has a scatter of at least $\pm 0.09$; (ix) mass of existing stars in region (from FADO fit). $^1$  This age differs greatly from the other regions and could be a significant overestimate if there is a problem with the oldest component fit.}
   \end{table*}
   
  \begin{table*}
   \begin{tabular}{lccccccccc}
   \hline
   Region & [OIII]4363 & [OIII]4959 & [OIII]5007 & [SII]6717 & [SII]6731 & $n_e$  &$\rm T_{e}$ &  $\rm 12+log10(O/H)$ \\ 
          & flux & flux & flux & flux & flux & $\rm cm^{-3}$ & K & `Amorin15'  \\
          \hline
    Nucleus  & $661(\pm 31)$ & 17028 & 51594 & 1003 & 742 & $62(\pm 20)$ & $12768 (\pm 228)$ & $8.08(\pm 0.02)$ \\
    Spur &  $21.3(\pm 2.3)$ & 638 & 1920 & 69.9 & 49.3 & $<40$ & $12703 (\pm 463)$ & $8.09(\pm0.04)$ \\
    Hotspot 1 &  $6.75(\pm 0.85)$ &  204 & 609 &    20.5 & 14.6 & $10(\pm 80)$ & $12078 (\pm 556)$ & $8.15(\pm 0.05)$\\ 
    Hotspot 2 & $2.97\pm 0.64)$ & 79 & 234 & 11.7 & 7.7 & low & $12689(\pm 1055)$ & $8.09(\pm 0.09)$ \\
    Redshifted region & - & 87 & 265 & 58.4 & 42.8 & $49(\pm 47)$ & - & -   \\
    Blueshifted region & $10.0(\pm 1.0)$ & 229 & 684 & 25.4 & 17.6 & $<49$ & $13430(\pm 545)$ & $8.02(\pm 0.05)$  \\
   Highest EW region & $23.7(\pm1.7)$ &  550 & 1649 & 69.6 & 47.6 & low & $13346(\pm 330)$ & $8.03(\pm 0.03)$ \\
    
   \hline
   \end{tabular}
    \caption{Line fluxes from region spectra in units of $10^{-16}$ erg $\rm cm^{-2}s^{-1}$ and the electron temperature $\rm T_e([OIII])$ and density $n_e$ derived from these fluxes using IRAF `temden'. Also the metallicity estimated from $\rm T_e$ via the relation of Amor\'in et al. 2015. [OIII]$\lambda$4363 is not detected for the 5th region.}
   \end{table*}

 NGC 4861, or at least the part covered by our data cube, seems relatively young, as our model fits (except for region 6, maybe) do not require stars older than 2.5 Gyr. On the other hand, the intermediate component  is over 1 Gyr old and this in combination with the high sSFR, 3.5 $\rm Gyr^{-1}$, and the star-formation histories (Fig 16, 17), is inconsistent with a constant SFR. Rather the SFR must have declined after formation of the disk, only for the galaxy to return to rapid star-formation more recently, producing the giant nebula, spur and $\rm H\alpha$ hotspots.

 The FADO fit to the nucleus ($r<6$ arcsec) spectrum estimates its stellar mass  at $1.38\times 10^7$ $\rm M_{\odot}$,  which with its current ($\rm H\alpha$)  SFR of 0.304 $\rm M_{\odot} yr^{-1}$ would have formed in 46 Myr. This is less than the mass-weighted mean age, given as 105 Myr and median $t_{1\over 2}$ age of 121 Myr. This implies that  (i) the initial strong starburst an estimated 125 Myr ago did not drive out all gas and halt further star-formation, (ii) the current $\rm H\alpha$ luminosity suggests that the SFR is increasing, or at a high point in a series of short bursts. This in turn could mean that gas is inflowing rather than being depleted and the giant nebula and `green pea' galaxy are growing.  
  
We propose that a low surface brightness disk galaxy (NGC 4861) formed 1 to 2 Gyr ago and more recently -- in the absence of evidence for interaction with another galaxy -- collided and merged with a massive HI cloud, and from $\sim 10^8$ years ago experienced a huge inflow of gas on its southern side which formed the high surface brightness giant nebula and a surrounding compact ($r_{exp}=0.5$ kpc) system of stars (Mrk 59). This morphology appears from some angles like a comet, but essentially the galaxy has entered a new stage  of evolution as an asymmetric `green pea' or `blueberry' starburst galaxy, which with the high gas content and extensive HI envelope seen in VLA observations  ($\sim1.1\times 10^9$ $\rm M_{\odot}$ of HI, $\rm M_{HI}>M_{stellar}$; Thuan, Hibbard \& L\'evrier 2004) might continue for another $\sim 10^8$ years or more.

Electron temperature $\rm T_e$ and electron density $n_e$ can be estimated together, using IRAF `temden' (Shaw \& Dufour 1994), with the flux ratio  of [OIII]$\lambda4363$ (where it is strong enough) to [OIII]$\lambda\lambda 4959,5007$ and the [SII]$\lambda 6717$ to [SII]$\lambda6731$ ratio. In Table 5 we give these fluxes (from FADO fit), firstly for the   $r<6$ arcsec nucleus region, where `temden' fits $n_e=\rm 62~cm^{-3}$ $(\pm 20)$ and $\rm T_e=12768~K$ $(\pm 228)$. These are  in the observed range for `green pea' galaxies with $n_e$ slightly lower than average (Izotov et al. 2011; Micheva et al. 2017). Outside the nucleus, $n_e$ measurements have large uncertainties and are generally lower (some are out of the fit range); the [SII] ratio becomes insensitive at $n_{e}<\rm 40~cm^{-3}$.

 Our data do not include [OII]$\lambda3727$ so we cannot use this to obtain a direct oxygen-abundance estimate. However,  the electron temperature based on [OIII]$\lambda 4363$ can be used to make another estimate of metallicity via the relation of Amor\'in et al. (2015) for `extreme emission line galaxies';  the central 
$\rm T_e([OIII])= 12768~K$ gives $\rm 12+log(O/H)=8.08$.  
We applied the same method to the other regions (Table 5), but their fainter [OIII]$\lambda4363$ lines give greater uncertainties.  We find the spur and hotspot 1 have similar $\rm T_e$ and metallicity to the central nebula and there are not great variations within the southern component. $\rm T_e$ and N2 estimators agree reasonably well. [OIII]$\lambda 4363$ emission could not be seen in our spectrum for the northern component (region 5), perhaps because of strong Balmer ($\rm H\gamma$) absorption. We had noted this region as having a slightly higher metallicity (8.20) on the basis of its higher  $\rm [NII]\lambda6584/H\alpha$, which may be evidence  of a (small) metallicity gradient.

  \onecolumn
  
\begin{figure}
\hskip -2.0cm
\includegraphics[width=1.35\hsize,angle=0]{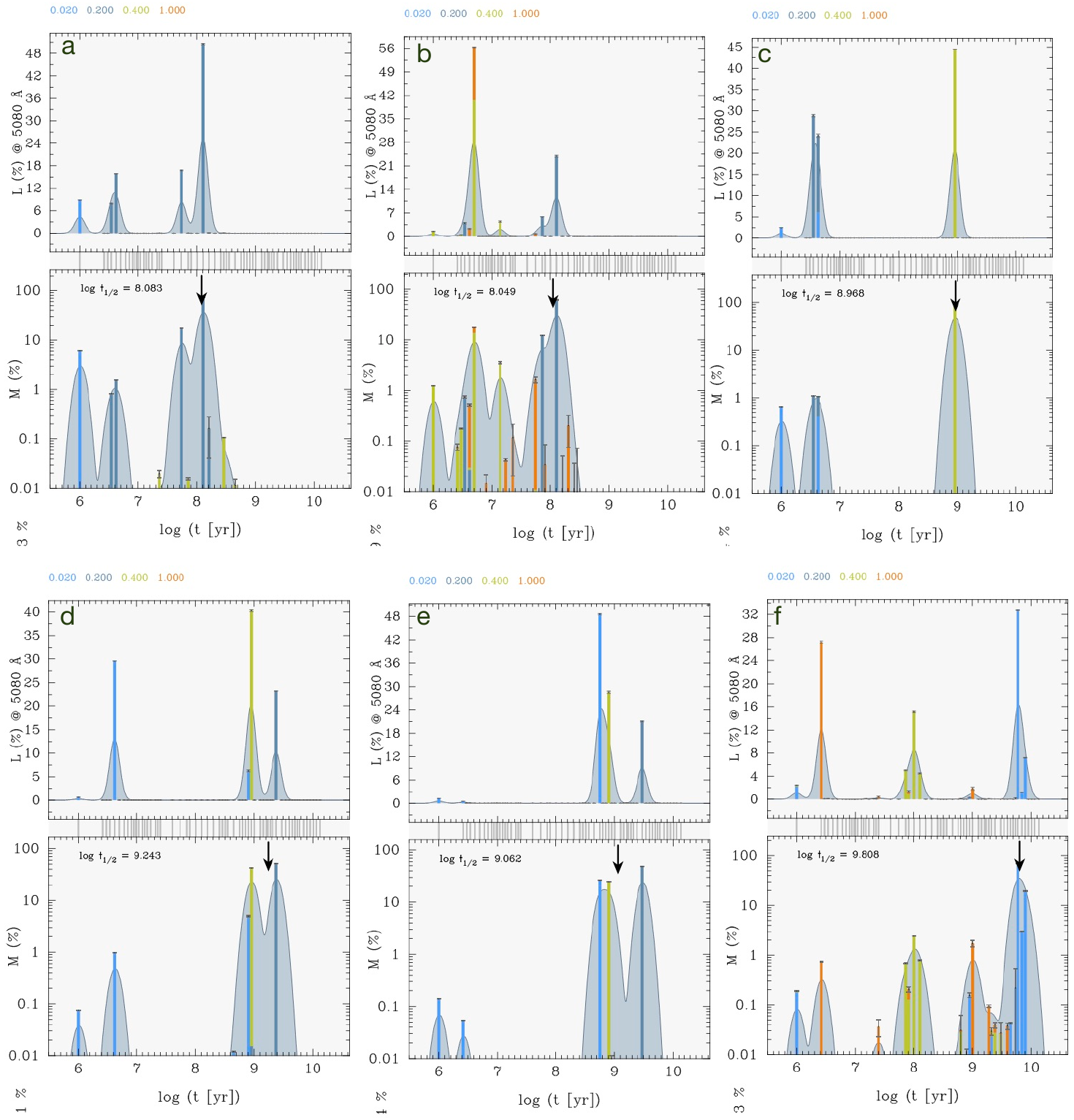}
\vskip -12.0cm
\caption{Distribution of stellar ages, weighting by luminosity (above, linear scale) and mass (below, log scale), as estimated by the FADO fit to the spectrum of these regions (the colours denote stellar metallicity relative to Solar): (top row, left to right) (a) nucleus, (b) spur , (c) hotspot 1; (below, left to right) (d) hotspot 2, (e) redshifted northern region (number 5), (f) blueshifted region (number 6). The vertical arrow in the mass fraction diagram marks the age when 50\% of the present-day stellar mass was formed.}
\end{figure}
%\twocolumn

\twocolumn

\begin{figure}
\vskip -2.0cm
\hskip  -2.5cm
\includegraphics[width=2.0\hsize,angle=0]{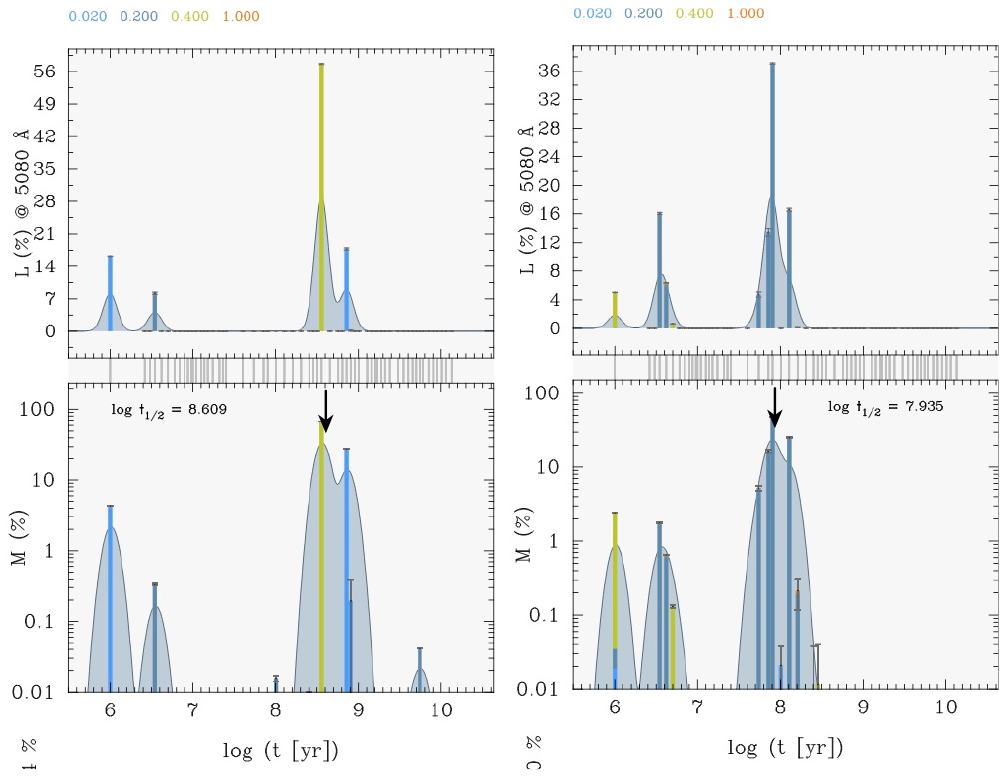}
\vskip -15cm
\caption{Distribution of stellar ages, weighting by luminosity (above, linear scale) and mass (below, log scale) , as estimated by the FADO fit to the spectrum of these regions: (left) region 7 selected as having $\rm EW>1000 \AA$,  (right) the whole galaxy, summed over 2114 spaxels.}
\end{figure}

     \section{Ionized Helium Line Emission}
   
   As originally discovered by Dinerstein \& Shields (1986) the NGC 4861 spectrum shows a double ionization helium emission line, HeII$\lambda 4686$, which is broadened to some extent, indicating an origin in Wolf-Rayet (WR) stars. The line map from FADO shows HeII$\lambda4686$ emission from the nucleus and (more faintly) the `Spur', but not elsewhere in the galaxy.   Summing HeII flux over central spaxels where it is above $0.4\times10^{-16}$ erg $\rm cm^{-2}s^{-1}arcsec^{-2}$, including the nucleus and spur, the total came to $1.95\times 10^{-14}$ erg $\rm cm^{-2}s^{-1}$.  The ratio of HeII$\lambda4686$ to $\rm H\beta$ integrated over the same area ($8.95\times 10^{-13}$ in 135 $\rm arcsec^2$) is 0.0218, about the same as in the WR galaxies (of similar metallicity) Mrk 71 (Micheva et al. 2017) and NGC 1569 (Mayya et al 2020).
   
  Due to the short, few Myr lifetime of massive O stars and their even briefer late phase as WR stars, the detection of WR features indicates a stellar population within a narrow age range, approximately 3--6 Myr (Mayya et al. 2020), although with binary star evolution taken into account this may extend to over 10 Myr (Eldridge \& Stanway 2022).   However, galaxies may also show narrow HeII$\lambda 4686$ emission lines originating from nebulae ionized by any source of sufficiently energetic photons ($h\nu >54.4$eV or 4 Rydberg), such as other types of extremely hot star (e.g. Sz\'ecsi ei al. 2015; Kehrig et al. 2015, 2018) or AGN (e.g. Wang \& Kron 2020).  
  
         \begin{figure}
 \hskip -1cm
 \includegraphics[width=1.2\hsize,angle=0]{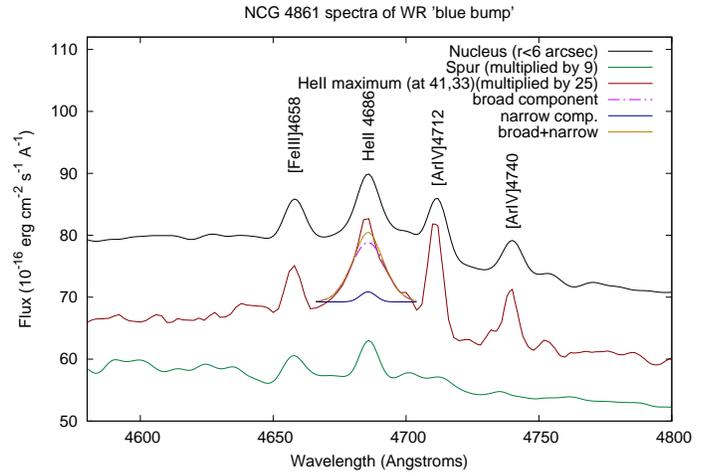}
 
\caption{The wavelength range containing the HeII$\lambda4686$ line and WR `blue bump', shown for the spectrum from the whole nucleus, the spur region, and from the pixel (41,33), within the nucleus, where the HeII line is strongest and most broadened (for this spectrum we show the broad and narrow components fitted to the HeII line). The latter two spectra are scaled upward for comparison with the first.} 
 \end{figure}

 \begin{figure}
 \vskip -1cm
    \hskip  -2.9cm
 \includegraphics[width=1.6\hsize,angle=0]{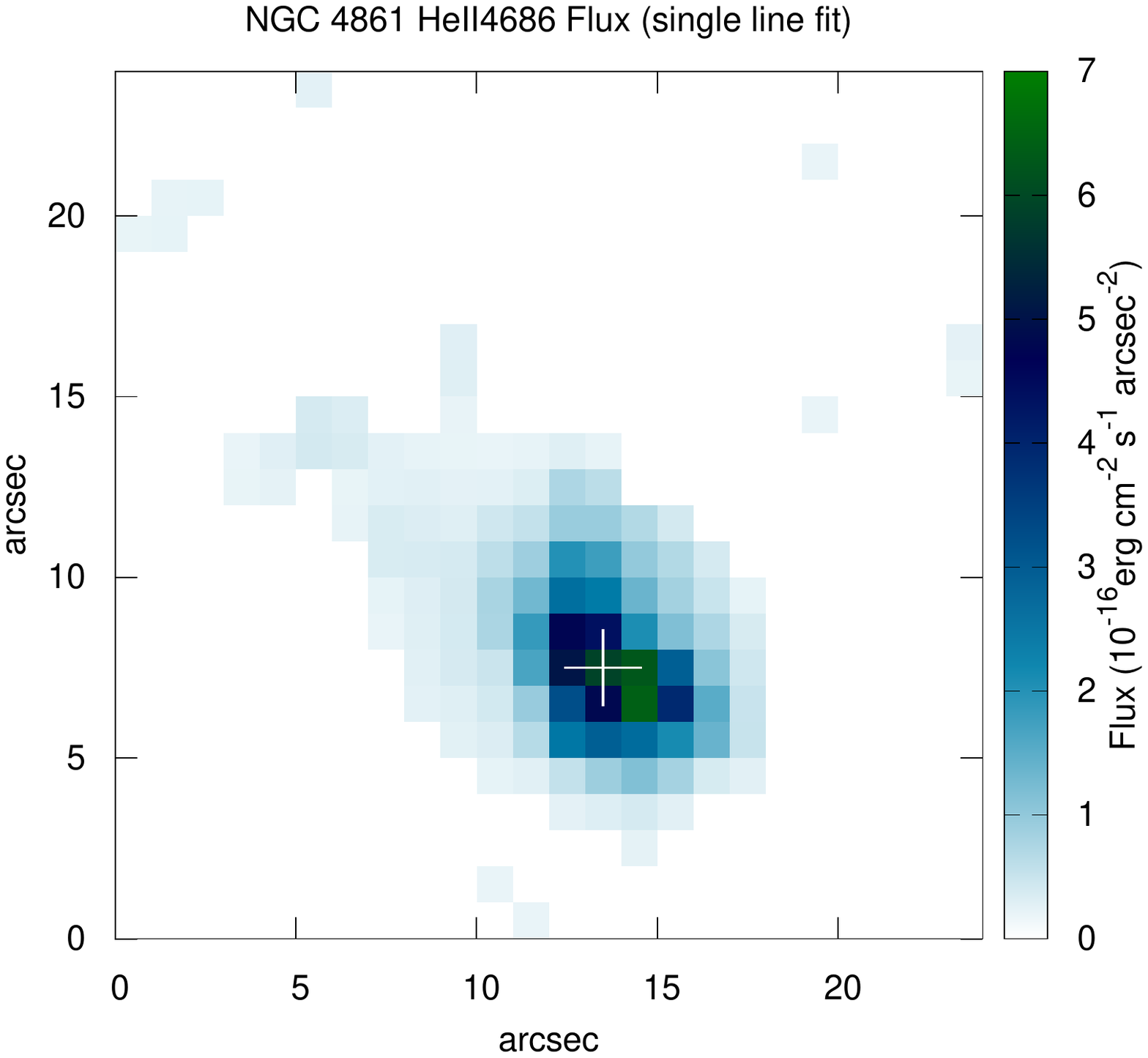}
 \vskip -1.9cm

 \hskip -2.9cm
 \includegraphics[width=1.6\hsize,angle=0]{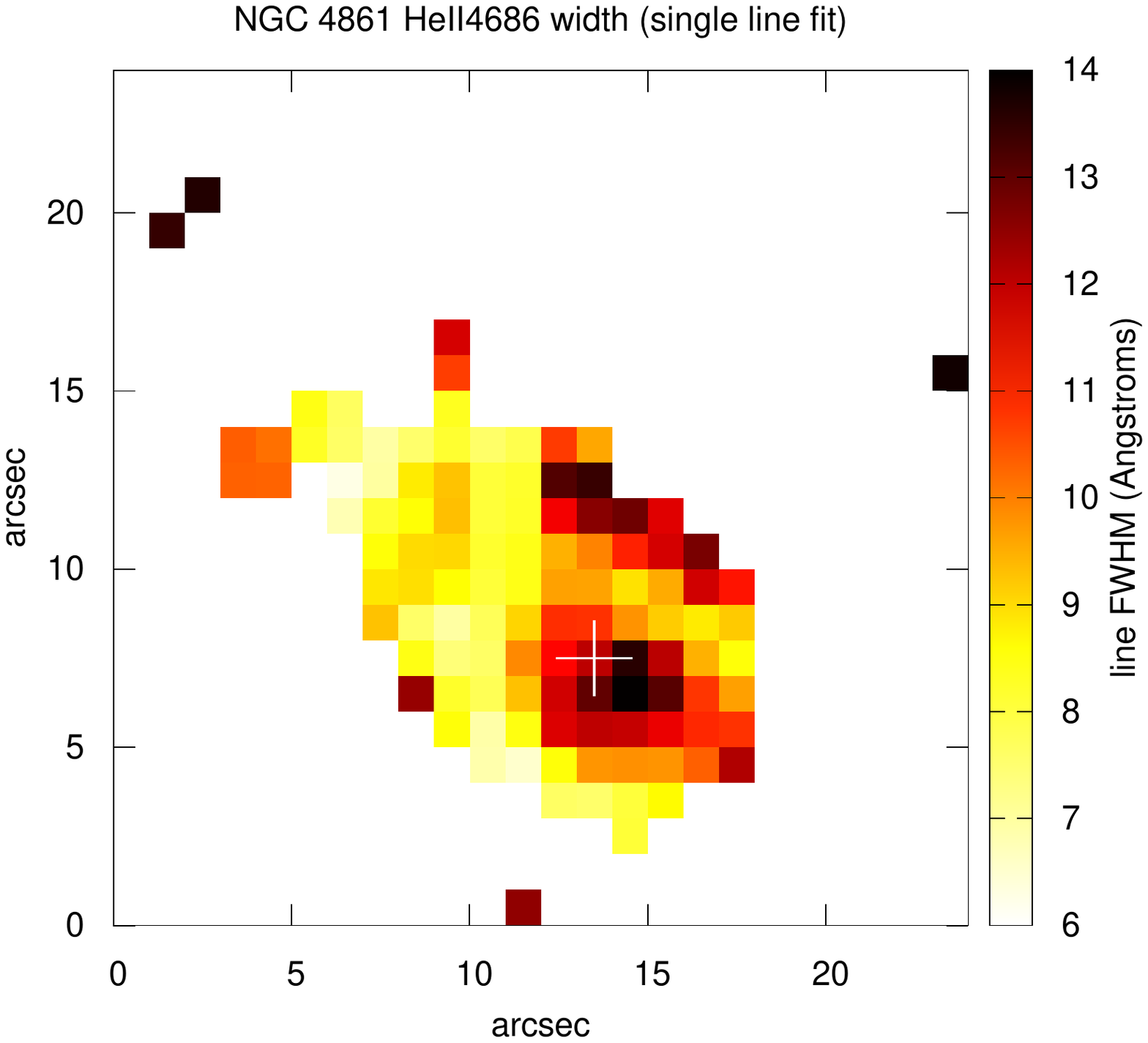}
\vskip -0.9cm
\caption{Map of Flux and Gaussian FWHM of HeII$\lambda4686$, from IRAF single-line fit (shown for spaxels with HeII flux above a threshold of 0.2 units of $10^{-16}$ erg $\rm cm^{-2}s^{-1}$)} 
 \end{figure}

   Fig 18 shows the spectrum around  HeII$\lambda4686$ for the nucleus and spur regions. In the first, HeII is clearly visible together with 3 narrow lines, which were  identified for this galaxy  as $\rm [FeIII]\lambda4658$ and $\rm [ArIV]\lambda\lambda4712,4740$ by Fernandes et al. (2004, Fig 7), and we confirm the observed line wavelengths fit. The HeII line has an EW of only $1.5\rm\AA$ but is visibly broadened, with $\rm FWHM\simeq 11\AA$ compared to $\rm 7\AA$ for other emission lines in this part of the spectrum (most of this from the instrumental resolution, $\rm FWHM\simeq 6\rm \AA$), implying that a substantial fraction of the emission is from WR stars. In the Spur region spectrum the HeII line is visible but weaker ($\rm EW \simeq 0.8\rm \AA$) and narrow ($\rm FWHM\simeq 7 \AA$), pointing to nebular emission rather than WR stars, and the two [ArIV] lines, which require a high energy (40.74eV), are very weak or absent.
     
  To further investigate the origin of the HeII emission we map both spatial distribution and line width.
  For this Section, `flux units' are defined as $10^{-16}$ erg $\rm cm^{-2}s^{-1}$.
   We fit Gaussian functions to the HeII line in individual spaxels using IRAF `fitprofile', over a $25\times25$ arcsec area enclosing the nucleus and spur. Firstly, a single Gaussian, fixing $\rm \lambda= 4685.71\AA$ and allowing the FWHM to vary freely.
 Fig 19 shows the flux and line width maps, with only spaxels with fitted fluxes  above 0.2 units   plotted (about $2\sigma$). Most of the HeII emission is from the nucleus, and there is also some from the  spur region with a lower surface brightness. The FWHM  is  7--$8\rm \AA$ in the spur and increases westwards, peaking at the nucleus,  with a maximum of $14\rm\AA$ at the spaxel (41,33), which is also the peak of the HeII flux. The HeII line from this spaxel is plotted in Fig 18, where it has the broadest wings.

    \begin{figure}
    \vskip -1.0cm
    \hskip  -2.9cm
 \includegraphics[width=1.6\hsize,angle=0]{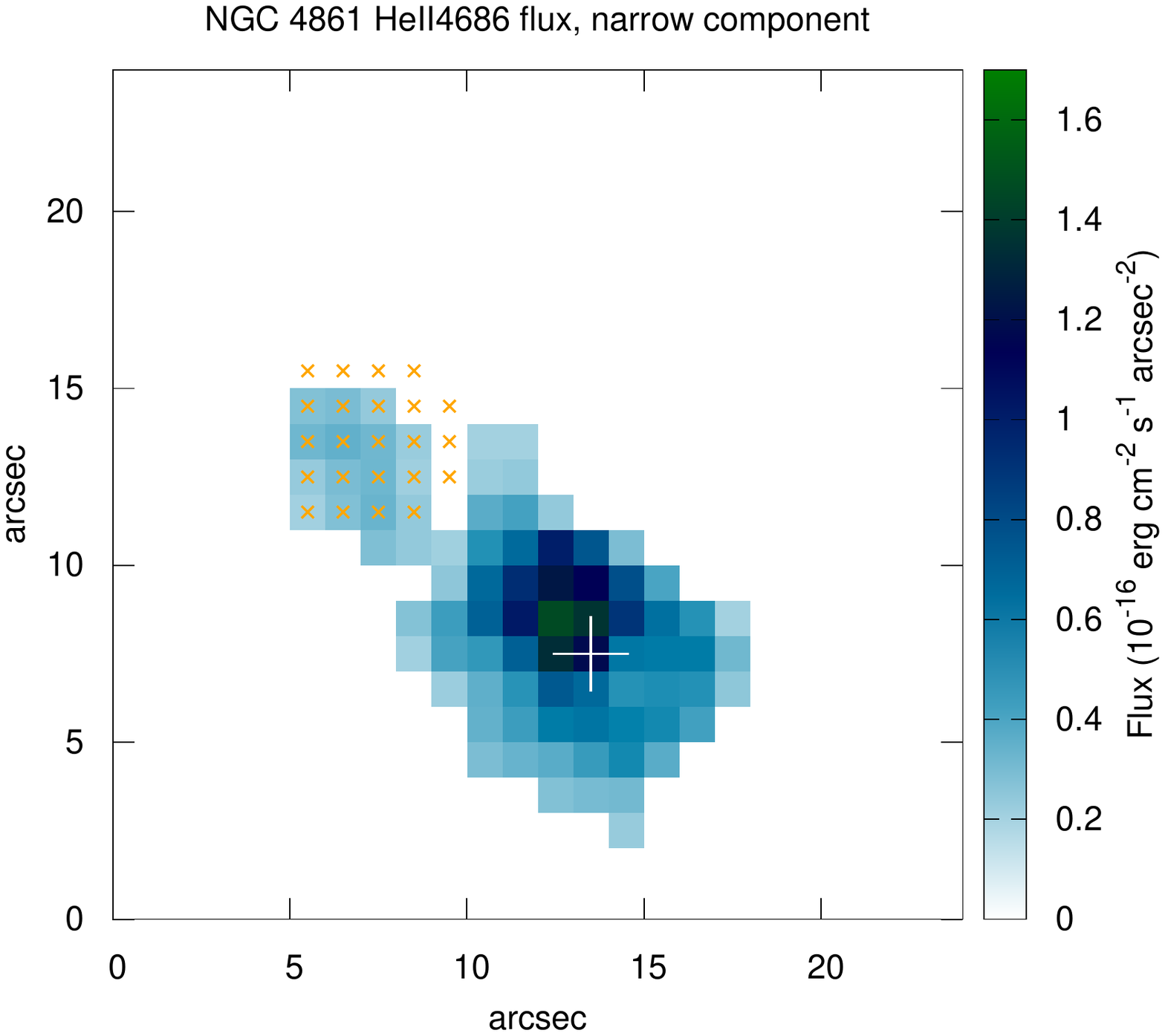}
 \vskip -1.9cm
 \end{figure}
 \begin{figure}
\hskip -2.9cm
 \includegraphics[width=1.6\hsize,angle=0]{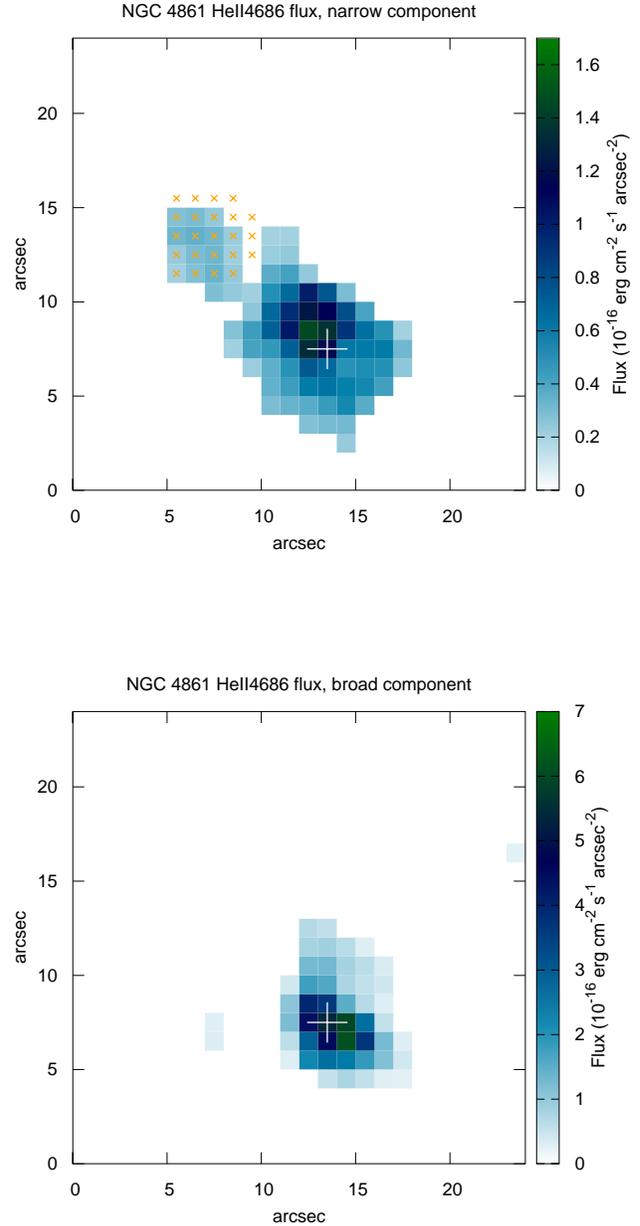}
\vskip -0.9cm
\caption{Map of Flux in HeII$\lambda4686$ line, separated into narrow ($7\rm \AA$) and broad ($15\rm \AA$) components by IRAF double Gaussian fits (only spaxels with flux $>0.2$ units plotted). The small orange `x' denote the Spur region as defined for the analysis of spectra, which is seen to be a region of significant (narrow) HeII emission.} 
 \end{figure}

 This spatial variation in FWHM suggests that there are two line components differing by a factor $\sim 2$ in width.  The broad component is likely to have about   $\rm FWHM\simeq 15\AA$, which is as observed for WNL type (late nitrogen rich) WR stars (Kehrig et al. 2013; Miralles-Caballero et al. 2016).
        We repeat our fitting with two Gaussians, both held at the same wavelength but with different  FWHM fixed at 7 and 15 $\rm\AA$, and allowed to vary in flux. Fig 20 shows the flux maps of the two fitted components. The broad component is concentrated in the central nebula, centred a little towards its western side, spans at most 9 arcsec (north to south), and is not seen elsewhere. The narrow component is lower surface brightness but more widely distributed, over the whole central nebula and spur, extending some 16 arcsec. The broad component flux is peaked at spaxel (41,33), flux 6.13 units, where we found the line to be broadest.  The narrow component is brightest  at spaxel (39,35), with flux  1.46. While they overlap, there is clearly some spatial offset between the broad (WR) and narrow (nebular) He II components.

  Some other WR galaxies show both broad and narrow HeII emission  components, with the narrow component being more spatially extended, e.g. Mrk 178 (Kehrig et al. 2013) and NGC 1569 (Mayya et al. 2020).   For NGC 4861,  the long-slit spectroscopy of Noeske et al. (2000) found a $\sim 1$ arcsec offset of the peak of the WR features from the peak of narrow HeII and $\rm H\beta$ emission, in the same direction of SW as we do. The WR stars might then be clustered slightly to the SW of the nebula centre, perhaps in the south-westerly of the three brightest spots in $\rm H\alpha$ (the other two are NE of centre), marked on the HST $\rm H\alpha$ image in Fig 21 (about 1.3 arcsec from the central point).  Miralles-Cabellero et al. (2016) examining generally larger galaxies in CALIFA noted WR regions were sometimes offset $\sim 1$ kpc from galaxy centres.
 Kehrig et al. (2013) attributed a 0.1 kpc offset between WR stars and nebular HeII in Mrk 178 to the effects of WR star winds.  This may be involved here -- however, the nebular HeII emission from the spur extends nearly 1 kpc from both the nucleus centre and the possible concentration of WRs, and could be powered by non-WR stars.

      \begin{figure}
\vskip -1.0cm
\hskip -3.6cm
  \includegraphics[width=1.8\hsize,angle=0]{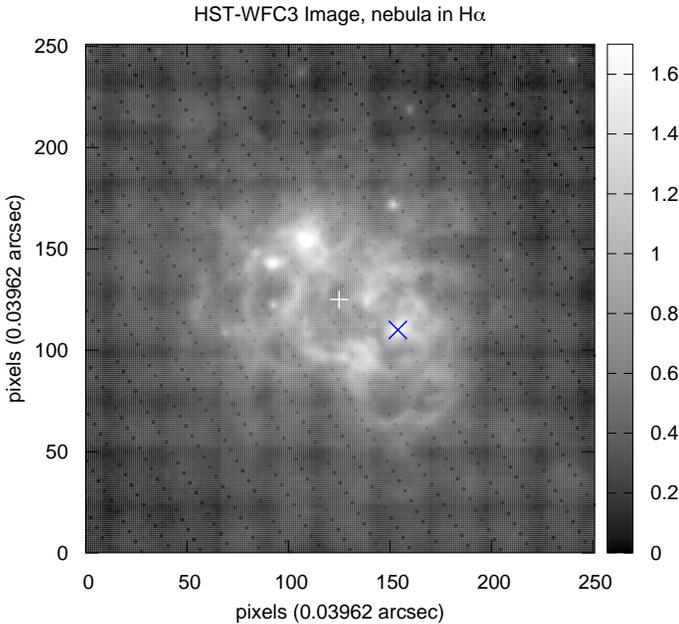}
  \vskip -1.2cm
\caption{Giant nebula shown in central $10\times10$ arcsec area of the f658N (showing $\rm H\alpha$), with a different scaling from Fig 2 so as to show features in the bright nucleus, with X marking proposed location of WR stars (RA 12:59:00.3, Dec +34:50:42).} 
 \end{figure}

 Summing the fitted line fluxes over the spaxels above a threshold 0.2 units gives 77.9 units  for the broad and
 40.4 for the narrow component (again of $10^{-16}$ erg $\rm cm^{-2}s^{-1}$). The narrow fraction is 34\%. Of the narrow component 
 4.68 units (12\%) comes from the spur region; interesting in that the spur has only 5\% the $\rm H\alpha$ or $\rm H\beta$ emission of the nucleus.
 These fluxes give the broad and narrow HeII luminosities as $2.36\times 10^{38}$ and $1.26\times 10^{38}$ erg $\rm s^{-1}$.
 
  Dividing this broad HeII luminosity by the mean WR star broad-component HeII luminosity given by Mayya et al. (2020), $1.22\times 10^{36}$ erg $\rm s^{-1}$, it is estimated to correspond to the emission of  192 WNL-type WR stars.  
   While this Gaussian line fitting is useful for studying spatial distributions, the fluxes are likely to be underestimates to some extent because of low signal/noise in many one-spaxel fits  and the overlap of HeII with other lines.
  The FADO fit gave an all-components   HeII$\lambda4686$ flux of $195\pm 4$ units, 1.65 times our (narrow plus broad total) 118.3. This made use of an observed spectrum with fitted stellar component (and nebular continuum) subtracted to leave only the emission lines, shown here on  Fig 22.
      \begin{figure}
\hskip -0.5 cm
  \includegraphics[width=1.1\hsize,angle=0]{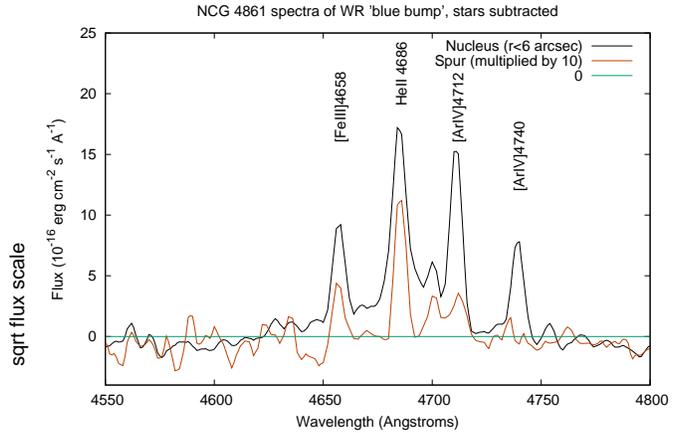}
  \vskip -0.5cm
\caption{Summed spectra of the nucleus and spur regions, in the vicinity of the WR `blue bump', with
stellar and nebular continua subtracted to leave only emission lines.} 
 \end{figure}
 After this subtraction, the nucleus spectrum shows a `blue bump' of above-zero values extending from about 4620 to $\rm 4750\AA$ and including the 4 emission lines already noted, and also a hint of  much broader underlying emission.  The Spur spectrum, with stars subtracted, shows two narrow emission lines and no broad component in this range.   From previous observations (e.g. Crowther \& Hadfield 2006) a very broad bump can be produced by some WRs with especially broad HeII plus further broad lines on its bluer side, including the triplets  NIII$\lambda\lambda\lambda$4634,4641,4642 and CIII$\lambda\lambda\lambda$4647,4650,4651, associated with WN-type WR stars and carbon-rich WC stars respectively.
 
 To investigate the possibility of a 3rd `extra broad' component, we  first sum the emission-lines-only datacube over 4620--$\rm 4750\AA$ to capture all of the emission in the `blue bump' range as an image (Fig 23). The emission region covers the whole central nebula with a small amount from the Spur. If we sum the fluxes in all spaxels above a threshold of 0.4 units (meaning the nucleus and spur), this comes to a surprisingly large  545.3 units.  

 Secondly, we perform multi-Gaussian fits to the spectrum of the central nebula with the stellar continua subtracted. We fit first the 4 narrow lines ($\rm FWHM=7\AA$) plus the broad ($\rm FWHM=15\AA$) HeII line. When this fit is subtracted from the spectrum, there is some sign of a residual bump above the zero-line with $\rm FWHM\sim 60\AA$. Also we can see a small, narrow line at  $\rm \lambda\simeq 4701\AA$, which is believed to be [FeIII]$\lambda$4701.6. 
 
 We then perform the fit again including a 6th component, a  $\rm FWHM=60\AA$ Gaussian fixed at $\rm 4650\AA$ to represent the CIII lines, and a 7th,  the extra [FeIII] line. This fit (Fig 24) gave the total fluxes of the narrow lines  (errors estimated by fitprof by repeatedly adding artificial background noise)
to be $56.0\pm 3.4$ for [FeIII]$\lambda4658$, $61.0\pm 5.2$ for narrow HeII$\lambda4686$,
  $121.2\pm 8.0$ for broad  HeII$\lambda4686$, $36.8\pm 2.3$ for [FeIII]$\lambda$4702, and  $110.8\pm 3.5$ and $53.1\pm 3.2$ for  [ArIV]$\lambda 4712$ and 4740. A very broad CIII component is fitted as 
  $87\pm 15$ units. The spur-region  spectrum shows the first two of these lines as significant detections, with fluxes  of $2.4\pm 0.7$ units for [FeIII]$\lambda4658$ and $7.5\pm 0.6$ for HeII.
The total HeII$\lambda$4686 flux of nucleus and spur from these fits is 189.7, consistent with the FADO estimate. The total of all 9 fitted line fluxes in the blue-bump comes to 526 units, near the total estimated from the summed image.

  \begin{figure}
  \vskip -1.4cm
\hskip -2.9cm
 \includegraphics[width=1.6\hsize,angle=0]{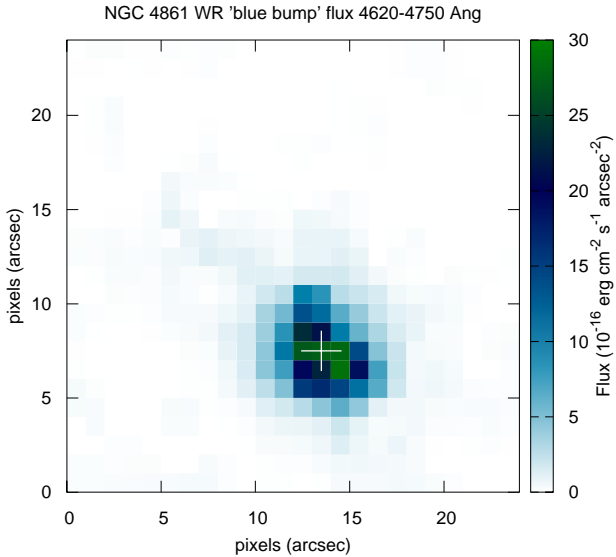}
\vskip -0.9cm
\caption{Map of Flux in the whole 4620--$\rm 4750\AA$ range, after subtraction of fitted stellar (and nebular) continuum, to leave the line emission including that from all types of WR stars and nebulae.} 
 \end{figure}
 \begin{figure}
 \hskip -0.9cm
   \includegraphics[width=1.1\hsize,angle=0]{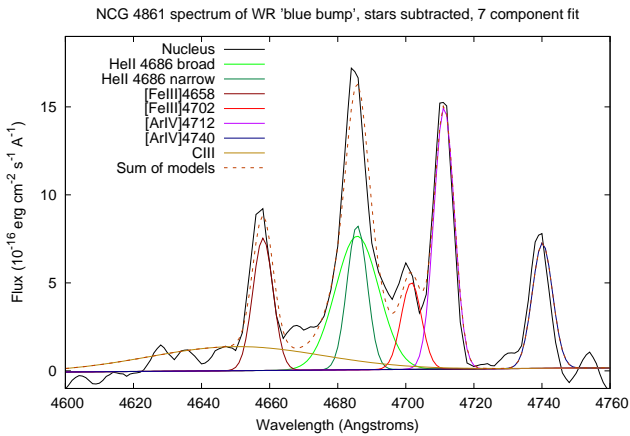}
  \vskip -0.9cm
\caption{Spectrum of the nucleus regions only (in the vicinity of the WR `blue bump'), with
stellar and nebular continua subtracted, compared to a fitted model with 6 emission line components including broad and narrow HeII and very broad CIII, shown separately and as their sum.} 
 \end{figure}

  NGC 4861, with both narrow and broad HeII components within a factor 2 in luminosity, and $\rm 12+log (O/H)\simeq 8.0$, lies in a mid-range (along with Mrk 178 and NGC 1569) between very low metallicity dwarf starburst galaxies like I Zw 18 and SBS0335-052E, where  HeII$\lambda4686$ emission is narrow (nebular) and strong (Kehrig et al. 2015, 2018), and large high-metallicity star-forming galaxies e.g. M83 (Hadfield et al. 2005) and a few galaxies in the CALIFA survey (Miralles-Caballero et al. 2016),  which contain many WCs as well as WN-type WR stars, giving broader  lines than the HeII seen here. This metallicity divide between narrow-line sources and WR stars as the dominant types of HeII emitter was seen in SDSS galaxies by Brinchmann, Kunth \& Durret (2008, Fig 14).  
  
  The FADO measurement of total HeII is probably an upper limit as it could include some flux from broader features  remaining after the stellar spectrum was subtracted,  whereas as noted above our fluxes from double-Gaussian fitting on top of the `bump' in single-spaxel spectra may be underestimated. The difference is a factor 1.65 so we could estimate the probable number of WN stars in the central nebula as in the range  192 to ($1.65\times 192$) 317. Another estimate is from the broad HeII flux in the 7-component fit to the nucleus spectrum, 121.2 ($\pm 8.0$) units, which corresponds to $3.68\times 10^{38}$ erg $\rm s^{-1}$, and 301 ($\pm 20$) WR stars.
    This is slightly higher than the number of WNL stars estimated for this galaxy (225) by Karthick et al. (2014), and lies between the estimates of Noeske et al. (2000) and Fernandes et al. (2004). 

  In addition, narrow (nebular)  emission makes up 36\% of the total  HeII$\lambda$4686 in these 
  fits, 68.5 units ($2.08\times 10^{38}$ erg $\rm s^{-1}$), and as we saw its sources are more widely distributed, over the whole of the nucleus and spur regions (though not detected elsewhere). The Spur emits about 7.5 units of this, and is a strong source of narrow HeII with its $\rm HeII/H\beta$ ratio coming to 0.018, almost equal to that of the nucleus from a narrow component alone. 
  
 Our 7-component fitting also suggested that a fraction of the `blue bump' flux  might come from a very broad feature such as CIII$\lambda 4650$ -- our fit estimated as $87\pm 15$ flux units, but the uncertainty will be greater because of uncertainty about its shape and the background level. The corresponding luminosity divided by the mean blended CIII/HeII luminosity of WC stars in the Large Magellanic Cloud from Crowther \& Hadfield (2006, table 2), $5\times 10^{36}$ erg $\rm s^{-1}$, gives an estimate of $\sim 53$ WC stars (in approximate agreement with Karthick et al. 2014). Our detection of broad and narrow HeII with different distribution is a much stronger result ($>10\sigma$), yet might require higher spectral/spatial resolutions to explain.
  
\section{Summary and Conclusions}
 
We obtained integral field spectroscopy of the nearby dwarf `cometary' starburst galaxy NGC 4861 using PMAS on the 3.5m Calar Alto telescope in Spain. Our pointing on the bright southern third of the galaxy contains a structure known as Markarian 59, centred on a high surface-brightness starbursting nucleus and giant HII nebula, and also includes one end of a lower surface brightness northern component, which is an edge-on disk galaxy. 
 Following earlier studies (e.g. Barth et al. 1994) we begin by mapping the $\rm H\alpha$ emission
 of the galaxy within our field of view, and we also compare with HST broad and narrow band imaging.
  From $\rm H\alpha$ emission we estimate 
a total SFR of 0.47--0.52 $\rm M_{\odot}yr^{-1}$ (in agreement with Karthick et al. 2014) . The greater part of this is taking place within  the 1 kpc diameter central nebula, but there is also active star-formation within an adjacent nebula we termed the `spur', two `hotspots' to the north, and elsewhere in the galaxy, including the northern component. The $\rm H\alpha$ equivalent width varies from 100$\rm \AA$ (northern component) to a peak at more than 1000$\rm\AA$, and the sSFR is estimated as at least 3.5 $\rm Gyr^{-1}$.

The excitation ratio $\rm [OIII]\lambda5007/H\beta$ is high over Mrk 59 and peaks at 7.35 in the giant nebula, a similar ratio to that in the starburst galaxies known as `green pea' or `blueberry' galaxies. This could be a combination of
the high star-formation density ($\rm 2~M_{\odot} yr^{-1}kpc^{-2}$ at the very centre), young age and the metallicity, low but not extremely low. There is a second, smaller and less high, peak in excitation which involves only the `hotspot 2'  within the northern component, where star-formation more than doubles the ratio.
The $\rm [NII]\lambda6584/H\alpha$  ratio remains low everywhere, pointing to a non-AGN galaxy with metallicity (estimated from strong line ratios) $\rm 12+log(O/H)\simeq 8.0$ in the central giant nebula increasing to 8.20 in the northern component (in agreement with e.g. Noeske et al. 2000, Fig 12). From [OIII]$\lambda\lambda$4363,5007 and [SII]$\lambda\lambda$6717,6731 line ratios we estimate electron temperature $\rm T_e=12768~K$ and density $\rm n_e=62~cm^{-3}$, and a similar metallicity.

We map the line-of-sight kinematics. There are some lower velocity motions (up to 40 km $\rm s^{-1}$) of the nebular regions in the southern component, and we select one region as being blueshifted. The visible part of the northern component is redshifted by $\sim 25$ km $\rm s^{-1}$ relative to the southern and this is probably due to the gradient in disk rotation, as seen by Thuan, Hibbard \& L\'evrier (2004). The galaxy does not seem be producing outflows above the escape velocity needed to expel gas (van Eymeren et al. 2009), and any of lower velocity will fall back as a `fountain'. As we discuss, the galaxy is more likely to be gaining HI.
       
We use the recently developed FADO spectral synthesis package to analyse more than 1000 spectra in the data cube, and attempt to reconstruct the spatially resolved star-formation history. This generated maps of luminosity-weighted  and mass-weighted mean stellar age, which 
appeared noisy. However, when the ages are averaged  over larger regions (of 16 to 115 spaxels here), a strong age gradient  emerges: youngest are the nucleus, spur and maximum EW regions (near the Mrk 59 centre), older are hotspot 1 and the blueshifted region, oldest are the hotspot 2 and the northern component.

The star-formation histories can be better time-resolved by running FADO on spectra summed over these regions. The nucleus and spur are seen to have formed in continuous bursts over the past 125 Myr and have a large component of $<10$ Myr age stars. The other regions have an intermediate-age population of $\sim 1$ to 2 Gyr old stars, plus a component of very young ($<10$ Myr) stars with its visible-light contribution ranging from 2\% for the northern galaxy `region 5' (which has a post-starburst spectrum), up to 55\% for hotspot 1 (with strong emission lines). Hotspot 2 is similarly a new starburst on top of a more massive intermediate-age population. This bimodal star-formation history resembles that fitted by Noeske et al. (2000), while  Amor\'in et al. (2012) fitted similar mixtures of young and older stellar populations for other `green pea' galaxies.

We also investigate the emission in $\rm HeII\lambda4686$, which requires very high energies and was detected from the giant nebula in previous studies. As first demonstrated in Kehrig et al. (2008), we show the power of IFS to investigate the HeII emission and WR content allowing us to present here a new view of the HeII origin in NGC 4861. 

The line FWHM varies with position, so that the HeII line could be considered as a mixture of narrow and broad ($\rm 15\AA$) components, which we quantify and map by fitting double-Gaussian profiles at the one-spaxel level. We find approximately two-thirds of the HeII flux is a broad component emitted from the centre of the giant nebula and could be explained by the presence of $\sim 300$ Wolf-Rayet stars of the Nitrogen type (confirming that much of the stellar content here formed less than 10 Myr ago). The remaining third is a narrow (nebular) component emitted from a more extended region --  all of the central nebula and the adjacent nebula we termed the Spur. Produced up to 1 kpc distant from the WR stars, this line may be emitted by other types of very hot stars associated with nebulae, but the source of ionization remains something of a mystery.
However, on the basis of previous studies, such a mixture of WR/broad HeII and other/narrow HeII may be as expected at this intermediate metallicity. There is weaker evidence for broader emission such as  $\rm CIII\lambda4650$, which if confirmed might be from Carbon-type WR stars.
 
 NGC 4861 appears to be a low surface brightness disk galaxy which formed relatively recently, 1 to 2 Gyr ago. The SFR then declined but reactivated $\sim 10^8$ years ago until the present day, with continuous starbursts, throughout the galaxy where it created hotspots, and especially at its southern end, where it formed a giant nebula and millions of stars. 
 This  major starburst event evolved the galaxy into an asymmetric  `green pea' or `blueberry' galaxy (depending on where these are divided by mass), or blue compact dwarf.
 
  The question is, how was this second burst triggered and fuelled when NGC 4861 appears not to be interacting or merging with any galaxy. VLA observations (Thuan, Hibbard  \& L\'evrier 2004, van Eymeren et al. 2009), however, show the galaxy  is very gas rich with an asymmetric HI envelope more extended than the stars ($\rm M_{HI}>10^9M_{\odot}$), and is also close to a star-free HI cloud, which could have been involved in triggering star-formation. As it is in such a gas-rich environment, maybe NGC 4861 already collided and merged with a large HI cloud resulting in gas inflow  on the south side of the disk.  This has been suggested for the formation of `cometary'/`tadpole' galaxies in general and could produce a lowered metallicity at the galaxy head (S\'anchez Almeida et al. 2013), or a more uniform composition due to rapid mixing (Lagos et al. 2016); NGC 4861 may be intermediate with 
 $\rm [NII]/H\alpha$ suggesting a small Z gradient, 0.2 dex upwards from the giant nebula to the disk.

\section*{Acknowledgments}

NR, JVM, JIP, CK, and SDP acknowledge financial support from the Spanish Ministerio de Econom\i'a y Competitividad under grant PID2019-107408GB-C44, from Junta de Andaluc\'ia under project P18-FR-2664, and also from the grant CEX2021-001131-S funded by
MCIN/AEI/ 10.13039/501100011033.

 PP thanks Funda\c c\~ao para a Ci\^encia e a Tecnologia (FCT) for managing research funds graciously provided to Portugal by the EU. This work was supported through FCT grants UID/FIS/04434/2019, UIDB/04434/2020, UIDP/04434/2020 and the project `Identifying the Earliest Supermassive Black Holes with ALMA (IdEaS with ALMA)' (PTDC/FIS-AST/29245/2017).

SDP is grateful to the Fonds de Recherche du Qu\'ebec - Nature et Technologies. SDP also acknowledges financial support from Juan de la Cierva Formaci\'on fellowship (FJC2021-047523-I) financed by MCIN/AEI/10.13039/501100011033 and by the European Union "NextGenerationEU"/PRTR.

This study is based on observations collected at the Centro Astron\'omico Hispano en Andaluc\'ia (CAHA) at Calar Alto, Spain, operated jointly by the Instituto de Astrof\'isica de Andaluc\'ia (CSIC) and Junta de Andaluc\'ia.  The CAHA Archive is part of the Spanish Virtual Observatory project 
funded by MCIN/AEI/10.13039/501100011033 through grant PID2020-112949GB-I00 CAB (INTA-CSIC). Part based on observations made with the NASA/ESA Hubble Space Telescope,  
and obtained from the Hubble Legacy Archive, which is a collaboration 
between the Space Telescope Science Institute (STScI/NASA), the Space 
Telescope European Coordinating Facility (ST-ECF/ESA) and the         
Canadian Astronomy Data Centre (CADC/NRC/CSA). This research has made use of the NASA/IPAC Extragalactic Database (NED), which is funded by the National Aeronautics and Space Administration and operated by the California Institute of Technology.

\section*{Data Availability}
The Calar Alto PMAS data can be found on the Calar Alto Archive at 
caha.sdc.cab.inta-csic.es/calto/. The data underlying this article will be shared on reasonable request to the corresponding author.
The HST-WFC3 data used here can be obtained from the MAST archive mast.stsci.edu, with Observation IDs
$\rm hst\_12497\_02\_wfc3\_uvis\_f814w\_ibse02$ and  
$\rm hst\_12497\_02\_wfc3\_uvis\_f658n\_ibse02$.
 
\section*{References} 

\vskip0.15cm \noindent Amor\'in R., P\'erez-Montero E.,  V\'ilchez J.M.,  Papaderos P., 2012,  ApJ, 749,185.

\vskip0.15cm \noindent Amor\'in R., et al., 2015, A\&A, 578, 105.

\vskip0.15cm \noindent Arp H., 1966, ApJS, 14, 1.

\vskip0.15cm \noindent Barth C.S., Cepa J., V\'ilchez J. M., Dottori H.A., 1994, AJ, 108, 2069. 

\vskip0.15cm \noindent Brinchmann J., Kunth D., Durret F., 2008, A\&A, 485, 657

\vskip0.15cm \noindent Breda I.,  V\'ilchez J.M.,  Papaderos P. ,  Cardoso L.,  Amor\'in R.O., 
 Arroyo-Polonio A.,  Iglesias-P\'aramo J.,  Kehrig C.,  P\'erez-Montero E., 2022, A\&A, 663, 29.

\vskip0.15cm \noindent Bruzual G., Charlot S., 2003, MNRAS, 344,1000.

\vskip0.15cm \noindent Cardamone C., et al., 2009, MNRAS, 399, 1191.

\vskip0.15cm \noindent Cardoso L.S.M., Gomes J.M., Papaderos P., 2019, A\&A. 622, 56.

\vskip0.15cm \noindent Chabrier G., 2003, PASP, 115, 763.

\vskip0.15cm \noindent Cid Fernandes R., et al., 2013. A\&A, 557, 86.

\vskip0.15cm \noindent Clarke L., et al., 2021, ApJ, 912, 22.

\vskip0.15cm \noindent Cochrane R.K., Best P.N., Sobral D., Smail I., Geach J.E., Stott J.P., Wake D. A., 2018, MNRAS 475, 3730. 

\vskip0.15cm \noindent Conti P.S., 1991, ApJ, 377, 115.

\vskip0.15cm \noindent Crowther P.A., Hadfield L.J., 2006, A\&A, 449, 711.

\vskip0.15cm \noindent Dinerstein H., Shields G., 1986, ApJ, 311, 45.

\vskip0.15cm \noindent Dottori H., Cepa J., V\'ilchez, J.M., Barth C.S., 1994, A\&A, 283, 753.

 \vskip0.15cm \noindent Eldridge J., Stanway E., 2022, arXiv:2202.01413

\vskip0.15cm \noindent van Eymeren J., Bomans D.J., Weis K., Dettmar R.-J., 2007, A\&A, 474, 67.

\vskip0.15cm \noindent van Eymeren J., Marcelin M., Koribalski B.S., Dettmar R.-J., Bomans D.J., Gach J.-L., Balard P., 2009, A\&A, 505, 105.

\vskip0.15cm \noindent Fern\'andez V., Amor\'in R., P\'erez-Montero E., Papaderos P., Kehrig C., V\'ilchez, J. M., 2022, MNRAS, 511, 2515.

\vskip0.15cm \noindent Fernandes, I. F., de Carvalho R., Contini T., Gal R.R., 2004, MNRAS, 355, 728.

\vskip0.15cm \noindent Gao Yu-Long, et al., 2017, RAA, 17, 41.
      
\vskip0.15cm \noindent  Garc\'ia-Benito R., et al., 2015, A\&A, 576, 135.
      
\vskip0.15cm \noindent Gomes J.M., Papaderos P.,  2017, A\&A, 603, 63.

\vskip0.15cm \noindent Guseva N.G.,  Papaderos P. ,  Izotov Y.I.,  Green R.F.,  Fricke K.J., Thuan T.X.,  Noeske K.G., 2003, A\&A, 407, 75.

\vskip0.15cm \noindent Hadfield L.J., Crowther P.A., Schild H., Schmutz W., 2005, A\&A, 439, 265.

\vskip0.15cm \noindent Henry A,., Scarlata C., Martin C.L.,  Erb D,  2015, ApJ, 809, 19.

\vskip0.15cm \noindent Izotov Y.I., Guseva N.G., Thuan T.X., 2011, ApJ, 728, 161

\vskip0.15cm \noindent Izotov Y.I., Guseva N.G., Fricke K.J., Henkel C., Schaerer D., Thuan T.X., 2021, A\&A, 646 138.

\vskip0.15cm \noindent Karthick M. C., L\'opez-S\'anchez, \'Angel R., Sahu D.K., Sanwal B.B., Bisht S., 2014, MNRAS, 439, 157.

\vskip0.15cm \noindent Kehrig C., V\'ilchez J.M., S\'anchez S.F., Telles E., P\'erez-Montero E.,  Mart\'in-Gord\'on, D., 2008, A\&A, 477,813.

\vskip0.15cm \noindent Kehrig C.,  P\'erez-Montero E.,  V\'ilchez J.M.,  Brinchmann J.,  Kunth D., Garc\'ia-Benito R., Crowther P., A.,  Hern\'andez-Fern\'andez J., Durret F., Contini T., Fern\'andez-Mart\'in A., James, B.L., 2013, MNRAS, 432, 2731.

\vskip0.15cm \noindent Kehrig C., V\'ilchez J.M., P\'erez-Montero E., Iglesias-P\'aramo J., Brinchmann J., Kunth D., Durret F., Bayo F. M, 2015, ApJ., 801L.

 \vskip0.15cm \noindent Kehrig C., V\'ilchez J. M., Guerrero M.A., Iglesias-P\'aramo J., Hunt L.K., Duarte Puertas S., Ramos-Larios G., 2018, MNRAS, 480,1081.

\vskip0.15cm \noindent Kennicutt Jr. R. C., 1998, ARAA, 36, 189.

\vskip0.15cm \noindent Lagos P.,  Demarco R.,  Papaderos P.,  Telles E.,  Nigoche-Netro, A., Humphrey A.,  Roche N.,  Gomes, J.M., 2016, MNRAS, 456,1549.

\vskip0.15cm \noindent Liu, Siqi, et al., 2022, ApJ, 927, 57.

\vskip0.15cm \noindent Marino R.A., et al.,  2013, A\&A, 559, 114.

\vskip0.15cm \noindent Mayya Y.D., et al., 2020, MNRAS, 498, 1496.

\vskip0.15cm \noindent Micheva G., Oey M.S., Jaskot A.E., James B.L., 2017, ApJ, 845, 165.

\vskip0.15cm \noindent Miralles-Caballero D., et al.., 2016, A\&A 592, 105. 

\vskip0.15cm \noindent Noeske K.G., Guseva N.G., Fricke, K.J., Izotov Y. I., Papaderos P., Thuan T. X., 2000, A\&A. 361, 33.

\vskip0.15cm \noindent Osterbrock, D.E. (ed.) 1989, Astrophysics of Gaseous Nebulae and Active
Galactic Nuclei (Mill Valley, CA: University Science Books)

\vskip0.15cm \noindent Paswan A.,  Saha K.,  Borgohain A.,  Leitherer C., Dhiwar S., 2022, ApJ, 929, 50. 

\vskip0.15cm \noindent Pappalardo C., et al., 2021, A\&A, 651, 99.

\vskip0.15cm \noindent Rela\~no M., Kennicutt R.C., Eldridge J.J., Lee J.C., Verlay S., 2012, 
MNRAS, 423, 2933. 

\vskip0.15cm \noindent Roche N., Humphrey A.  Gomes J.M., Papaderos P., Lagos P., S\'anchez S.F.,
2015, MNRAS, 453, 2349.

\vskip0.15cm \noindent Roth M.M., et al., 2005, PASP,117, 620

 \vskip0.15cm \noindent S\'anchez S.F., et al., 2012, A\&A, 538, 8.

\vskip0.15cm \noindent S\'anchez Almeida J., Mu\~noz-Tu\~n\'on C., Elmegreen D.M., Elmegreen B.G., M\'endez-Abreu J., 2013, ApJ, 767, 74. 

\vskip0.15cm \noindent Schaerer D., Contini T., Pindao M., 1999, A\&AS, 136, 35.

\vskip0.15cm \noindent Schlafly E., Finkbeiner D., 2011, ApJ, 737, 103.

\vskip0.15cm \noindent Shaw R.A., Dufour R.J., 1994, ASPC, 61, 327.  

\vskip0.15cm \noindent Sz\'ecsi D., et al., 2015, A\&A, 581, 15. 

\vskip0.15cm \noindent Thuan T.X., Hibbard J.E., L\'evrier F., 2004, AJ, 128, 617.

\vskip0.15cm \noindent Wakamatsu K. -I., Sakka K., Nishida M., Jugaku J., 1979, PASJ, 31, 635.

\vskip0.15cm \noindent Wang Q., Kron R.G., 2020, MNRAS, 498, 4550.

\vskip0.15cm \noindent Wild V., et al., 2014, A\&A, 567, 132. 

 \end{document}